\pdfoutput=1
\RequirePackage{ifpdf}
\ifpdf 
\documentclass[pdftex]{sigma}
\else
\documentclass{sigma}
\fi

\numberwithin{equation}{section}

\newtheorem*{Theorem*}{Theorem}

 { \theoremstyle{definition}

 }

\begin{document}
\allowdisplaybreaks

\newcommand{\arXivNumber}{2109.13900}

\renewcommand{\PaperNumber}{045}

\FirstPageHeading

\ShortArticleName{Notes on Worldsheet-Like Variables for Cluster Configuration Spaces}

\ArticleName{Notes on Worldsheet-Like Variables\\ for Cluster Configuration Spaces}

\Author{Song HE~$^{\rm abcde}$, Yihong WANG~$^{\rm abdf}$, Yong ZHANG~$^{\rm gh}$ and Peng ZHAO~$^{\rm b}$}

\AuthorNameForHeading{S.~He, Y.~Wang, Y.~Zhang and P.~Zhao}

\Address{$^{\rm a)}$~School of Fundamental Physics and Mathematical Sciences,\\
\hphantom{$^{\rm a)}$}~Hangzhou Institute for Advanced Study, UCAS, 310024 Hangzhou, P.R.~China}
\EmailD{\href{mailto:songhe@itp.ac.cn}{songhe@itp.ac.cn}, \href{mailto:yihongwang141@ucas.ac.cn}{yihongwang141@ucas.ac.cn}}

\Address{$^{\rm b)}$~CAS Key Laboratory of Theoretical Physics, Institute of Theoretical Physics,\\
\hphantom{$^{\rm b)}$}~Chinese Academy of Sciences, 100190 Beijing, P.R.~China}
\EmailD{\href{mailto:pzhao@tjufz.org.cn}{pzhao@tjufz.org.cn}}

\Address{$^{\rm c)}$~International Centre for Theoretical Physics Asia-Pacific, Beijing/Hangzhou, P.R.~China}

\Address{$^{\rm d)}$~School of Physical Sciences, University of Chinese Academy of Sciences,\\
\hphantom{$^{\rm db)}$}~No.~19A Yuquan Road, 100049 Beijing, P.R.~China}

\Address{$^{\rm e)}$~Peng Huanwu Center for Fundamental Theory, Hefei, 230026 Anhui, P.R.~China}

\Address{$^{\rm f)}$~Laboratoire d'Annecy-le-Vieux de Physique Th\'eorique, Universit\'e Savoie Mont Blanc,\\
\hphantom{$^{\rm f)}$}~9~Chemin de Bellevue, 74941 Annecy-le-Vieux, France}

\Address{$^{\rm g)}$~Department of Physics and Astronomy, Uppsala University, 75108 Uppsala, Sweden}

\Address{$^{\rm h)}$~Perimeter Institute, 31~Caroline Street North, Waterloo, Ontario,  N2L 2Y5, Canada}
\EmailD{\href{mailto:yzhang@perimeterinstitute.ca}{yzhang@perimeterinstitute.ca}}

\ArticleDates{Received November 21, 2022, in final form June 29, 2023; Published online July 12, 2023}

\Abstract{We continue the exploration of various appearances of cluster algebras in scattering amplitudes and related topics in physics. The cluster configuration spaces generalize the familiar moduli space ${\mathcal M}_{0,n}$ to finite-type cluster algebras. We study worldsheet-like variables, which for classical types have also appeared in the study of the symbol alphabet of Feynman integrals. We provide a systematic derivation of these variables from $Y$-systems, which allows us to express the dihedral coordinates in terms of them and to write the corresponding cluster string integrals in compact forms. We mainly focus on the $D_n$ type and show how to reach the boundaries of the configuration space, and write the saddle-point equations in terms of these variables. Moreover, these variables make it easier to study various topological properties of the space using a finite-field method. We propose conjectures about quasi-polynomial point count, dimensions of cohomology, and the number of saddle points for the $D_n$ space up to $n=10$, which greatly extend earlier results.}


\Classification{13F60; 05E14; 81T30}

\section{Introduction and review}

In~\cite{Arkani-Hamed:2020tuz,Arkani-Hamed:2019plo}, {\it cluster configuration spaces} have been introduced, which generalize the moduli space ${\mathcal M}_{0,n}$ (corresponding to the type-$A_{n-3}$ space) to cases corresponding to other finite-type cluster algebras. The idea behind these spaces is that of {\it stringy canonical forms}~\cite{Arkani-Hamed:2019mrd}, which are Euler--Mellin type integrals providing $\alpha'$ deformations of {\it canonical forms}~\cite{Arkani-Hamed:2017tmz} of any polytopes. The prototypes of such string-like integrals are genus-zero (open- and closed-) string integrals, which are very special stringy canonical forms for the so-called Arkani-Hamed--Bai--He--Yan (ABHY) associahedra describing tree amplitudes in bi-adjoint $\phi^3$ theory
as $\alpha' \to 0$~\cite{Arkani-Hamed:2017mur}. For generalized associahedra of any finite-type cluster algebra, one can define a class of special (real and complex) integrals called {\it cluster string integrals} whose $\alpha'\to 0$ limit gives the canonical forms of their ABHY realizations, which are rigid and natural extensions of the usual string integrals. The corresponding cluster configuration space is defined by compactifying the integration domain with a boundary structure mirroring that of the generalized associahedron, which manifests ``factorizations'' of the integrals at finite $\alpha'$ dictated by the Dynkin diagram~\cite{Arkani-Hamed:2019plo}; the positive space is literally a~curvy generalized associahedron polytope. For type $A_{n{-}3}$, we recover the Deligne--Knudsen--Mumford compactification of ${\mathcal M}_{0,n}$~\cite{deligne1969irreducibility, Knudsen_1983} and in general the spaces become generalizations of the latter. As we will review shortly, a cluster configuration space is defined in terms of a~set of constrained variables that are related to $Y$-systems~\cite{Zamolodchikov:1991et}, and it is natural to wonder if one could find certain unconstrained variables as generalizations of worldsheet variables of ${\mathcal M}_{0,n}$ (type $A$). In this note, we will show that it is indeed the case, and such worldsheet-like variables become very useful for our study of cluster string integrals, the ABHY polytopes, and the corresponding saddle-point equations. In particular, we will use these variables for a systematic study of topological properties, e.g., the number of saddle points (or Euler characteristic), for such configuration spaces, especially for type $D_n$~\cite{Arkani-Hamed:2020tuz}.

On the other hand, the authors of~\cite{Chicherin:2020umh} have recently proposed such variables for classical types $ABCD$, which naturally appear as symbol letters for Feynman integrals and scattering amplitudes in QFTs. They were obtained from clever but somewhat ad hoc birational maps of the kinematical variables.
A goal of the current paper is to provide a \emph{systematic} derivation of such polynomials from the corresponding $Y$-systems.
We find simple birational transformations from $d$ initial $y$-variables to $d$ worldsheet-like variables $z_1, z_2, \dots, z_d$, and all other $y$-variables can be obtained from $Y$-systems; by our transformation they generate polynomials of $z$'s. As we will show, there are natural transformations such that these polynomials of $z$'s take particularly simple forms, including the familiar $z_i-z_j$ factors for the type-$A$ moduli space. In terms of these variables, we can alternatively define the cluster configuration space as the $d$-dimensional space with all $N$ polynomials of $z$'s non-vanishing. We will illustrate how this works with the most important example, the configuration space of type $D$. We also write down the alphabet for $E_6$ and the non-simply-laced types, leaving the technical details to an upcoming work~\cite{WangZhao}. These worldsheet-like variables play a crucial role for two purposes: they naturally appear as letters of the symbol for certain classes of Feynman integrals (especially for type $D_n$ in various ladder integrals to all loops~\cite{He:2021mme}), and they are the natural variables for describing the cluster configuration spaces for which we can define cluster string integrals, saddle-point equations,~etc.\looseness=1

In addition to providing a systematic derivation of such $z$-variables, we will also apply these variables to the study of cluster string integrals and important topological properties of cluster configuration spaces, such as the number of saddle points. In~\cite{Arkani-Hamed:2020tuz}, it has been shown that the number of saddle points for types $A_d$, $B_d$ and $C_d$ is $d!$, $d^d$ and $(2 d)!!/2$, respectively. This was done by computing the number of points in the configuration space for a finite field $\mathbb{F}_p$ (with a generic prime $p$), which turns out to be a degree-$d$ {\it polynomial} of $p$ for these cases; by plugging in $p=1$ we obtain the Euler characteristic, or the number of saddle points up to a~sign. In these cases, coefficients of the polynomial give dimensions of cohomology, which can be generated by independent ${\rm d}\log$ forms.\footnote{Similar computations using the finite-field method has played an important role in determining the number of saddle points etc.\ for the so-called Cachazo--Early--Guevara--Mizera (CEGM) amplitudes~\cite{Cachazo:2019ngv}, which are closely related to stringy integrals of Grassmannian case~\cite{Arkani-Hamed:2019rds}; see~\cite{Sturmfels:2020mpv} and especially the appendix of~\cite{Agostini:2021rze} for such results in the context of likelihood equations.} Although such computations can also be done with, e.g., $F$-polynomials, at least for types $A$ and $C$, the use of worldsheet variables has been crucial to obtaining general result (for type $B$ it is coincidental that a new set of ``linear'' variables can be used to show that the space is the complement of a hyperplane arrangement, similar to type~$A$). In~\cite{Arkani-Hamed:2020tuz}, it has been conjectured that the point count for types $D_4$, $D_5$ and $G_2$ are {\it quasi-polynomials}, i.e., polynomials with correction terms that depend on the characteristic $p$. The point count also nicely gives Euler characteristics for these spaces, but the lack of such worldsheet variables for type $D_n$ has made computations much more difficult. We will partially solve this problem by performing such a point count for $D_n$ up to $n=10$, which is unimaginable without the help of these variables. Moreover, we are also interested in linear relations among~${\rm d}\log$ forms made of such polynomials especially for type $D_n$, which will be instrumental to ``bootstrap'' all possible integrable symbols with such an alphabet~\cite{He:2021mme}.

Before proceeding, here we will first give a lightening review of the cluster configuration space and associated string integrals, and refer the readers to~\cite{Arkani-Hamed:2020tuz,Arkani-Hamed:2019mrd} for details. It turns out that a~gauge-invariant way for describing the compactified moduli space, which allows us to see all the boundaries of the $A_{n-3}$ configuration space explicitly, is by introducing $N:=n(n{-}3)/2$ dihedral coordinates $u_{a,b}$ (one for each diagonal of an $n$-gon labeled by $(a,b)$). They satisfy the same number of constraints called {\it $u$-equations}~\cite{Brown:2009qja,Koba:1969kh,Koba:1969rw}:
\[
 1-u_{a,b}=\prod_{(c,d)~{\rm incompatible~with}~(a,b)} u_{c,d},
\]
where on the right-hand side we have the product over all the $u_{c,d}$ variables where $(c,d)$ is incompatible with $(a,b)$ (the diagonals cross). It is a remarkable fact that these $n(n{-}3)/2$ equations have a solution space of dimension $n{-}3$, which we call the open $A_{n{-}3}$ configuration space for all $u_{a,b} \neq 0$. In the real case, it is referred to as the positive part if we further ask all $u_{a,b}>0$. The $u$-equations have revealed that the configuration space is a {\it binary geometry}, which has boundary structures that ``factorize'' by removing a node of the (type-$A$) Dynkin diagram: as any $u_{c,d}\to 0$, all incompatible $u_{a,b}\to 1$, thus the space factorizes as $A_L \times A_R$ where~$L$,~$R$ are associated with the two polygons of the $n$-gon separated by $(c,d)$. For the positive (or more precisely, non-negative) part, all the variables $0\leq u_{a,b}\leq 1$ and the space has the shape of a (curvy) associahedron. In an analogous manner, one introduces {\it $u$-variables} and {\it $u$-equations} for any finite-type cluster algebra: for any Dynkin diagram $\Gamma$ with $d$ nodes (the rank of the cluster algebra), we have $N_\Gamma$ such variables and equations with $N_\Gamma$ the dimension of the cluster algebra.\footnote{For $A_d$, $N_\Gamma=d(d{+}3)/2$, for $B_d$ or $C_d$, $N_\Gamma=d(d{+}1)$, for $D_d$, $N_\Gamma=d^2$, and for $E_6$, $E_7$, $E_8$, $F_4$ and $G_2$, we have $N_\Gamma=42, 70, 128, 28$ and $8$, respectively.} Explicitly, we have
\[
1-u_I=\prod_{J=1}^{N_\Gamma} u_J^{J | I}
\]
for all $I=1,2,\dots, N_\Gamma$, where $J | I$ is the so-called {\it compatibility degree} from cluster variable $J$ to~$I$~\cite{Arkani-Hamed:2019vag,Fomin:2001rc}. Beyond type $A$, these degrees go beyond $0$, $1$ and we have $J|I>0$ if and only if the two cluster variables are incompatible (only for simply-laced cases, we have $I|J=J|I$). Remarkably, the $N_\Gamma$ equations for $N_\Gamma$ variables again have an $d$-dim solution space where $d$ is the rank of the cluster algebra, which is called the {\it cluster configuration space} ${\mathcal M}_\Gamma$ of type $\Gamma$. As any $u_J \to 0$, all incompatible $u_I \to 1$ (those with $J | I>0$), and for the positive part (all $u$ between $0$ and $1$), it has the shape of the corresponding generalized associahedron with $N_\Gamma$ facets. Note that our $u$-variables are related to $y$-variables of $Y$-systems via $u_I:=y_I/(1+y_I)$ for $I=1, 2, \dots, N_\Gamma$, and it is a non-trivial fact that the $u$-equations become equivalent to recurrence relations of the corresponding $Y$-system. One way to parameterize ${\mathcal M}_\Gamma$ is to use $d$ initial $y$-variables $y_1, \dots, y_d$ (which correspond to an {\it acyclic} quiver), and by solving $Y$-systems the remaining $N_\Gamma-d$ $y$'s are rational functions of them. Equivalently, all the $y$- or $u$-variables can be expressed as ratios of $F$-polynomials and monomials of $y_1, \dots, y_d$. The positive part, ${\mathcal M}^+_\Gamma$, where we have all $0<u_I<1$ or all $y_I>0$, is thus parameterized by $y_i>0$ ($i=1, 2, \dots, d$) for any initial acyclic quiver. As we have mentioned, it is more advantageous to use unconstrained variables, e.g., for explicitly computing such string integrals (either as an expansion in $\alpha'$ or even at finite $\alpha'$), or studying their saddle-point equations. However, we emphasize that the $u$-variables constrained by the $u$-equations provide a {\it gauge-invariant} description of the configuration space.

Associated with such a space, it is natural to write the cluster string integral, e.g., for the positive part ${\mathcal M}_\Gamma^+$~\cite{Arkani-Hamed:2019mrd}
\begin{gather}
I_\Gamma(\{X\}):=\big(\alpha'\big)^d \int_{{\mathcal M}^+_{\Gamma}} \Omega({\mathcal M}^+_\Gamma)~\prod_{I=1}^{N_\Gamma} u_I^{\alpha' X_I} ,
\label{stringyintegral}
\end{gather}
where the integral is over ${\mathcal M}^+_\Gamma$ (with all $u_I>0$), and we do not need the explicit expression of the canonical form $\Omega\big({\mathcal M}^+_\Gamma\big)$ but the property that on any boundary $u_I \to 0$, it becomes the canonical form of the latter which takes a ``factorized'' form by removing a node of the Dynkin diagram; the integral is ``regulated'' by the ``Koba--Nielsen'' factor with exponents $X_I>0$. This is exactly the domain of convergence, which also cuts out an ABHY generalized associahedron~${\mathcal A}_\Gamma$ with each facet reached by $X_I \to 0$. As $\alpha'\to 0$, the leading order of the integral is given by the {\it canonical function} of the ${\mathcal A}_\Gamma$, and most remarkably even for finite $\alpha'$, the integral has ``perfect'' factorization as each $X_I \to 0$~\cite{Arkani-Hamed:2019plo}.

In particular, for type $A_{n{-}3}$ we have the following $n$-point string integral over (the positive part of) the configuration space
\[
I_{A_{n{-}3}} (\{X\}):=\big(\alpha'\big)^{n{-}3} \int_{{\mathcal M}^+_{0,n}} \Omega\big({\mathcal M}^+_{0,n}\big)~\prod_{(a,b)} u_{a,b}^{\alpha' X_{a, b}} ,
\]
where we have alternatively denoted the space of type $A_{n{-}3}$ as the moduli space ${\mathcal M}_{0,n}$, and~$X_{a,b}$ can be identified with the {\it planar variables}, one for each facet of the ABHY associahedron~${\mathcal A}_{n{-}3}$. The $\alpha'\to 0$ (low-energy) limit of this open-string integral gives the canonical function of ${\mathcal A}_{n{-}3}$, which is nothing but the (diagonal) bi-adjoint $\phi^3$ tree amplitude~\cite{Arkani-Hamed:2017mur}. These physical string integrals, as a special case of cluster string integrals, still factorize at $X_{a,b}=0$, even at finite $\alpha'$, which nicely reflects the ``perfect'' factorization of ${\mathcal M}^+_{0,n}$ as one can see from the $u$-equations. Now let us quickly see how ${\mathcal M}_{0,n}$ and the worldsheet variables naturally appear from the cluster configuration space of type $A_{n{-}3}$. The $u$-variables as defined above do not refer to any worldsheet picture, but the $u$-equations can be nicely solved once we introduce worldsheet variables $z_i$ for $i=1,2,\dots, n$ with an ${\rm SL}(2)$ redundancy. The $u_{i,j}$'s are exactly the dihedral coordinates written as cross ratios of the marked points on the boundary:
\begin{gather}
u_{i,j}=\frac{z_{i-1,j} z_{i,j-1}}{z_{i-1,j-1} z_{i,j}},\qquad 1\leq j<i-1<n ,
\label{crossratio}
\end{gather}
where $z_{i,j}:=z_j -z_i$, and it is easy to see that the $z$-variables provide the solution to the $u$-equations, though we still need to fix the ${\rm SL}(2)$ redundancy, e.g., by choosing $z_1=-1$, $z_2=0$, $z_n=\infty$. The open cluster configuration space can be identified with the moduli space ${\mathcal M}_{0,n}$ and in particular for the positive part, we have $z_i$'s ordered, $z_1<z_2<\cdots<z_n$ (in our fixing, we have $0<z_3<\cdots<z_{n{-}1}$. One can rewrite $I_{A_{n{-}3}}$ in terms of the $z$-variables once we realize that $\Omega\big({\mathcal M}^+_{0,n}\big)$ is given by the Parke--Taylor form:
\[
I_{A_{n{-}3}}:= (\alpha' )^{n{-}3}~\int_{{\mathcal M}_{0,n}^+} \frac{{\rm d}^{n{-}3} z}{z_{1,2}z_{2,3} \cdots z_{n, 1}} \prod_{j<i} |z_{i, j}|^{\alpha' s_{i, j}},
\]
where the measure is the top, ($n-3$)-form on ${\mathcal M}_{0,n}$, and we have rewritten the factor $\prod u_{a,b}^{X_{a,b}}$ as the Koba--Nielsen factor with Mandelstam variables $s_{i,j}:=(k_i+k_j)^2=2 k_i \cdot k_j$ (subject to momentum conservation). It is easy to see that linearly-independent Mandelstam variables can be written as
\[
s_{i,j}=X_{i+1,j} + X_{i,j+1} - X_{i,j} - X_{i+1,j+1}.
\]
More precisely, in the gauge-fixing above the factor reads $\prod_{1\leq j<i\leq n-1} z_{i,j}^{\alpha' s_{i,j}}$,
where we have the same number, $n(n{-}3)/2$, of linear factors (since $z_{1,2}=1$ drops out). We will refer to the collection of the $n(n{-}3)/2$ polynomials of degree $1$ as an {\it alphabet}; in our gauge-fixing it consists of $\{z_i, 1+z_i\}$ for $i=3,\dots, n{-}1$ and $\{z_{i,j}\}$ for $3\leq j<i\leq n{-}1$. The cluster configuration space is equivalent to the complement of the hyperplane arrangement in terms of the letters of the alphabet, i.e., $\{ z_{i,j}=0, z_i=0, 1+z_i=0\}$, and one can easily derive various topological properties from here, e.g., the number of saddle points is $(n{-}3)!$~\cite{Arkani-Hamed:2020tuz}. Moreover, the number of connected components of ${\mathcal M}_{0,n}(\mathbb{R})$ is $(n{-}1)!/2$, and they are given by all possible orderings of $z_3, z_4, \dots, z_{n{-}1}$ (or sign patterns of the alphabet), which are in $1-1$ correspondence with all the $(n{-}1)!/2$ consistent sign patterns of $u$-variables~\cite{Arkani-Hamed:2020tuz}.

In the following, we will systematically introduce worldsheet-like variables for cluster configuration spaces of $D_n$ type, and study properties of the space and string integrals. We sketch how the construction can be generalized for other finite types and present results for the alphabet for the classical types and $E_6$, $F_4$, and $G_2$, leaving technical details to a separate work~\cite{WangZhao}.

In Section~\ref{sec2}, we will show how to derive from $Y$-systems {\it worldsheet-like variables} $z_1, \dots, z_d$, which are related to the initial $y_1, \dots, y_d$ via birational maps, such that all the $N_\Gamma$ $u$ (or $y$) variables become ratios of polynomials of $z$'s. As we have seen above, naively there are more than $N_\Gamma$ polynomials in $z$'s, but similar to the type-$A$ case, there is a natural ${\rm SL}(2)$ gauge redundancy. After a choice of gauge, we find exactly $N_\Gamma$ such polynomials ({\it letters}), and we call their collection an {\it alphabet}. Nicely, we find linear factors similar to those in type $A$, $z_i$, $1+z_i$ and~$z_{i,j}$, and polynomials of higher degrees: for types $B$, $C$ and $D$, there are quadratic polynomials, and for types $E_6$, $F_4$ and $G_2$, there are polynomials of degree at most $4$, $5$ and $4$, respectively.\footnote{
The form of an alphabet depends on a choice of gauge. In this work, we find that the same canonical gauge choice as the type-$A$ case produces the alphabets in the literature. More discussion on gauge choices will be given in~\cite{WangZhao}.}
We will use type $D_n$ as the main example: we obtain nice formulae that express $u$-variables as ``cross ratios'' of the letters and discuss how the boundaries of ${\mathcal M}^+_{D_n}$ can be obtained by degeneration of these $z$-variables.

In Section~\ref{sec3}, we proceed to write cluster string integrals and saddle-point equations using such $z$-variables. The Koba--Nielsen factor present in the case of $A_n$ cluster algebras naturally generalizes for the other cluster string integrals as the product of all letters. Moreover, we focus on the positive configuration space, whose canonical form may be written in terms of the $z$-variables and is analogous to the Parke--Taylor factor. We will see how factorization on ``massless'' poles of stringy integral correspond to boundaries of the positive part of the configuration space as $z$-variables pinch (as shown explicitly for, e.g., $D_4$, $B_3$ and~$G_2$). Using $z$-variables, we will write the $D_n$ scattering equations which also provide a diffeomorphism from the positive configuration space to the ABHY polytope. Last but not least, we will numerically solve these equations, and count the number of solutions (saddle points) up~to~$D_7$.

In Section~\ref{sec4}, we shall study the topological properties of the cluster configuration spaces using the worldsheet variables. By counting points in the complement of the hypersurfaces defined by the letters over a finite field, we obtain quasi-polynomials that encode information on the Euler characteristic and the dimensions of cohomology up to $D_{10}$. Moreover, we will conjecture the number of independent $2$- and $3$-forms for any $D_n$, which are not affected by the ``correction terms'' in the quasi-polynomials. We provide a further support for the conjecture by explicitly writing down the cohomology relations among the 2-forms that provide an upper bound on the number of independent 2-forms.

\section{Worldsheet-like variables for finite-type cluster algebras}\label{sec2}

The cluster configuration space ${\mathcal M}_{\Gamma}$ is the configuration
space of $u$-variables, which are initially defined in~\cite{Arkani-Hamed:2020tuz, Arkani-Hamed:2019plo} as
the solution of the $u$-equations over the complex numbers:
\begin{gather}
1-u_I=\prod_{J=1}^{N_\Gamma} u_J^{J | I} ,
\label{global}
\end{gather}
where the compatibility degree $J|I$ is a non-negative integer assigned to each pair of the $u$-vari\-ables. The possible values for $J|I$ vary for different underlying cluster types of the $u$-variables. This ``global'' form of the $u$-equations is a natural generalization of the relations among the dihedral coordinates of a disc with $n$ points on the boundary.
\begin{figure}
\centering
\includegraphics[scale=0.45]{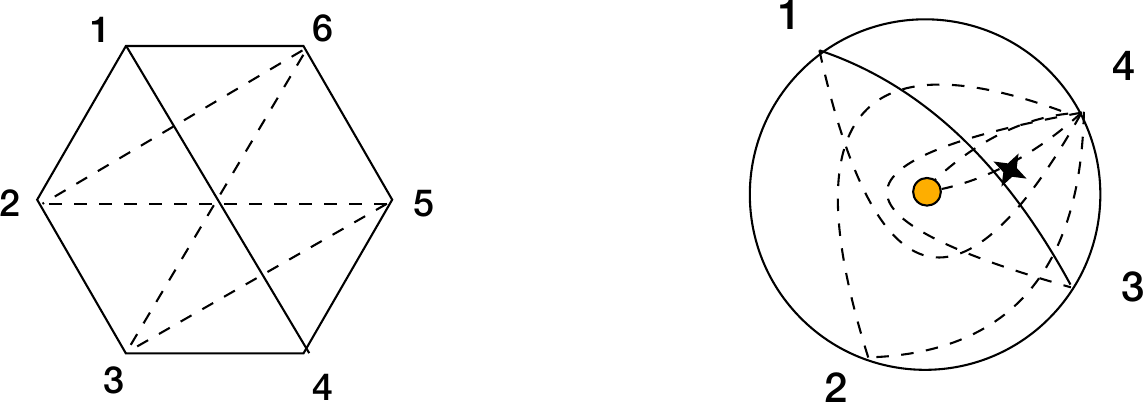}
\caption{(1) The $A_{n-3}$ worldsheet is a disc with $n$ boundary points. For each diagonal, the incompatible diagonals are the ones that cross it. (2) The $D_n$ worldsheet is a once-punctured disc with $n$ boundary points. The diagonals may go around the puncture in two orientations. There are tagged and untagged arcs connecting boundary points to the interior puncture.}
\end{figure}
For example, the $A_3$ $u$-equations are
\begin{gather*}
1-u_{1,3} = u_{2,4} u_{2,5} u_{2,6} ,\qquad
1-u_{1,4} = u_{2,5} u_{2,6} u_{3,5} u_{3,6} ,
\end{gather*}
and cyclic permutations of the indices. The $u$-variables are the cross ratios~\eqref{crossratio} of the marked points.

The $D_n$ moduli space is another distinguished example that admits a geometric realization as a once-punctured disc with $n$ boundary points~\cite{Fomin_2008}. In addition to the arcs $u_{i,j}$ connecting two boundary points there are two arcs $u_i$, $\tilde u_i$ connecting a boundary point to the interior puncture, one untagged and one tagged. Note that unlike in the $A$ type, there are nontrivial arcs $u_{i,i+1}$ connecting two neighboring boundary points that go around the puncture. Moreover $u_{j,i}$ and~$u_{i,j}$ are different arcs that go around the puncture in the opposite orientations. For example, the~$D_4$ $u$-equations are
\begin{gather*}
1-u_{1,2} = u_3 \tilde{u}_3 u_4 \tilde{u}_4 u_{2,3} u_{2,4} u_{3,4}^2 u_{3,1} u_{4,1} , \\
1-u_{1,3} = u_4 \tilde{u}_4 u_{4,1} u_{4,2} u_{2,4} u_{3,4} , \\
1-u_1 = \tilde{u}_2 \tilde{u}_3 \tilde{u}_4 u_{2,3} u_{2,4} u_{3,4} , \\
1-\tilde u_1 = {u}_2 {u}_3 {u}_4 u_{2,3} u_{2,4} u_{3,4} ,
\end{gather*}
and cyclic permutations of the indices. It is not known whether the $u$-variables can be written as ``cross ratios'' of the worldsheet coordinates, which is the main problem that we will address in this section.

While the $E_n$ types are not known to admit geometric realizations, their $u$-equations may be abstractly defined as~\eqref{global}. The equations for the non-simply-laced types may be written by ``folding'' the arcs and setting equal a subset of the $u$-variables~\cite{Arkani-Hamed:2019plo}.

The definition of the global $u$-equations~\eqref{global} is equivalent to a recursive definition by the local $u$-equations~\cite{Arkani-Hamed:2020tuz}
\begin{gather}
\left(\frac{1-u_{v}}{u_{v}}\right)\left(\frac{1-u_{v'}}{u_{v'}}\right)=\prod_{w\ne v}(1-u_{w})^{a_{v,w}} .\label{eq:locu}
\end{gather}
Here $a_{v,w}$ is the Cartan matrix for the Dynkin diagram $\Gamma$. One chooses an acyclic quiver whose graph is $\Gamma$ and assigns an initial set of $u$-variables to the $d$ nodes. Since an acyclic quiver has at least a source and a sink, we perform a mutation on a source or a sink that reverses the directions of the arrows. A new $u_{v'}$ is generated by~\eqref{eq:locu}. The full set of $N_\Gamma$ $u$-variables is generated by successive applications of the mutation.

The local $u$-equations make it possible for us to connect the $u$-variables with well-studied objects in the cluster algebra literature as (\ref{eq:locu}) is the defining relation for Zamolodchikov's $Y$-systems under the map $y_{v}=u_{v}/(1-u_{v})$ and the $y$-variables are subsets of cluster variables~\cite{Fomin:2001rc}, also known as cluster $\mathcal{X}$-coordinates~\cite{Fock:2003xxy}. The integers $J|I$ are identified with the compatibility degrees between two corresponding elements $y_{I}$, $y_{J}$ in $Y$-systems~\cite{Zamolodchikov:1991et}.

\begin{figure}[t]
\centering
\includegraphics[scale=0.22]{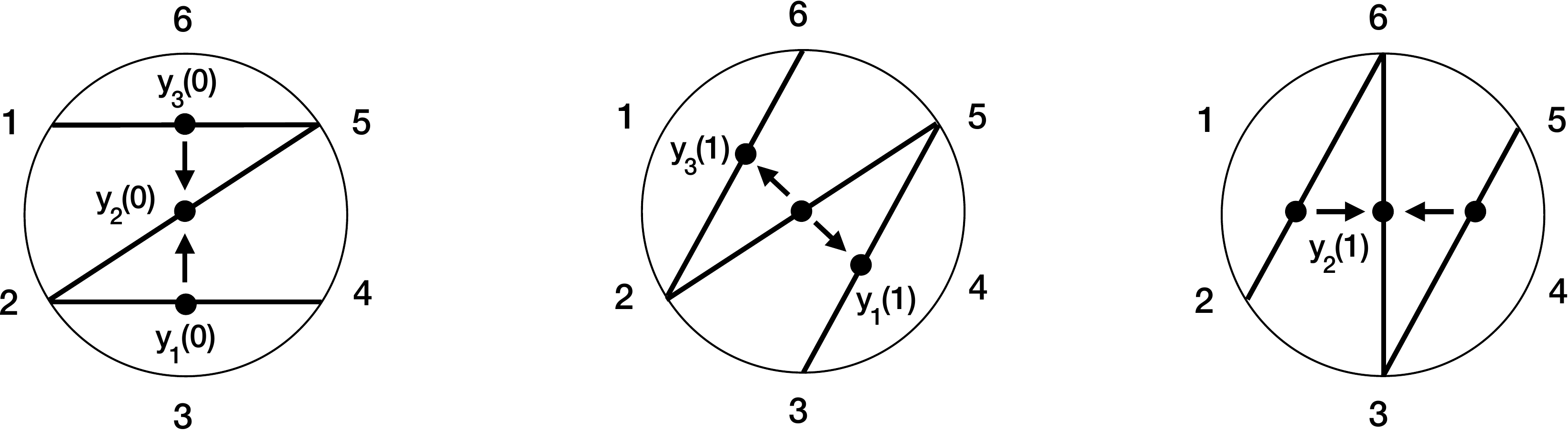}
\caption{The $A_{3}$ worldsheet. Each diagonal in a triangulation corresponds to a node in the quiver and is assigned a $y$-variable. We consider an initial zigzag triangulation where each node is either a source or a sink. At each step, we mutate the source nodes (or equivalently, flip the diagonals) and generate new $y$-variables according to the $Y$-system equations.}
\label{AnWS}
\end{figure}

Let us begin by reviewing how the local $u$-equations or $Y$-system equations arise from the~$A_{n-3}$ configuration space, which is the moduli space of a disc with $n$ points on the boundary. Consider a triangulation of the $n$-gon by non-crossing diagonals. To each $(i,j)$ diagonal we associate a $u_{i,j}$ (or $y_{i,j}$)-variable. Each diagonal is contained in some quadrilateral. We obtain another triangulation by a \emph{flip}: replacing a diagonal with another in the quadrilateral, as shown in Figure \ref{AnWS}. The local $u$-equations express relations between diagonals that are related by a flip
\begin{gather}
\frac{1-u_{i,j}}{u_{i,j}} \frac{1-u_{i+1,j+1}}{u_{i+1,j+1}} = (1-u_{i,j+1})(1-u_{i+1,j}) .
\label{localu}
\end{gather}
The local $u$-equations~\eqref{localu} take particularly simple forms in terms of the $y$-variables
\[
y_{i,j} y_{i+1,j+1} = (1+y_{i,j+1})(1+y_{i+1,j}) .
\]

One can associate a quiver with a triangulation where each diagonal corresponds to a node and there is an arrow if one diagonal follows another in the counterclockwise direction around a~common vertex~\cite{Gekhtman_2005}. Consider a quiver where each node is either a source or a sink. It gives rise to a zigzag triangulation, as shown in Figure \ref{AnWS}. There is a notion of time evolution where each step is represented by mutations on all the source nodes or on all the sink nodes~\cite{Fomin_2003}. This is equivalent to a simultaneous flip on the corresponding diagonals. We choose to mutate on the source nodes. The diagonals in the $(i,i+1,j,j+1)$ quadrilateral will be mapped as
\[
y_{i,j} \to y_{i+1, j+1}.
\]

Using the label in terms of the nodes of the Dynkin diagram and the number of simultaneous mutations performed on the sources or the sinks, the local $u$-equations are equivalent to the $Y$-system equations
\begin{gather}
y_{i}(t-1) y_i(t) = \prod_{j \to i} (1 + y_j(t))^{-a_{i,j}} \prod_{i \to j} (1 + y_j(t-1))^{-a_{i,j}} .
\label{Y-system}
\end{gather}
We will use $y_{i}(t)$, $u_{i}(t)$ to label the Dynkin node and time step, and $y_{i,j}$, $u_{i,j}$ to denote the polygon label.
A key property of $Y$-systems of Dynkin type is periodicity: the $y$-variables return to their initial values after a finite number of mutations. The period divides $2(h+2)$, where $h$ is the Coxeter number~\cite{Zamolodchikov:1991et}. The cross-ratio representation of $y$-variables was used to prove the periodicity conjecture for the $A_n$ type~\cite{Volkov_2007}.

For the $D_n$ type, there are additional equations for the extra nodes in the Dynkin diagram~\cite{Arkani-Hamed:2020tuz}
\begin{gather}
\frac{1-u_{i-1,i}}{u_{i-1,i}} \frac{1-u_{i,i+1}}{u_{i,i+1}} = (1-u_{i-1,i+1}) (1-u_{i})(1-\widetilde u_{i}) , \nonumber\\
\frac{1-u_{i}}{u_{i}}\frac{1-\widetilde u_{i+1}}{\widetilde u_{i+1}} = 1-u_{i,i+1} , \nonumber\\
\frac{1-\widetilde u_{i}}{\widetilde u_{i}}\frac{1- u_{i+1}}{ u_{i+1}} = 1-u_{i,i+1} .
\label{localD}
\end{gather}
Our main result is a solution of the $D$-type equations in terms of a set of generalized cross ratios of worldsheet $z$-variables
\begin{alignat}{3}
&u_{i, j}=\frac{z_{i,j-1} z_{i-1,j}}{z_{i,j} z_{i-1,j-1} }, \qquad && u_{j, i} = \frac{w_{i,j-1} w_{i-1,j}}{w_{i,j} w_{i-1,j-1}} ,& \nonumber \\
&u_{i}=\frac{z_{i,n+3} w_{i-1,i} }{z_{i-1,n+3} w_{i,i}}, \qquad &&\widetilde u_{{i}} =\frac{z_{i,n+2} w_{i-1,i}}{z_{i-1,n+2} w_{i,i}},&
\label{main}
\end{alignat}
where $1 < j < i \le n+1$. When $i, j = 1$, the variables are defined cyclically, e.g., $u_{1,j} := u_{n+1, j}$ and $u_{1} := u_{n+1}$.
Here $w_{i,j}$ is a cubic polynomial of the form
\[
w_{i,j} = z_{1,n+3} z_{i,j} z_{n+1,n+2} - z_{1,n+1} z_{i,n+3} z_{j,n+2}.
\]
The next section is devoted to a derivation of this result. We explicitly derive the formulae for the $D_4$ type, which allow us to infer the general form. One can then verify that the generalized cross ratios solve the $D_n$ $u$-equations for any $n$.

\subsection[The gluing construction of the D\_n configuration space]{The gluing construction of the $\boldsymbol{D_n}$ configuration space}

\begin{figure}
\centering
\includegraphics[scale=0.24]{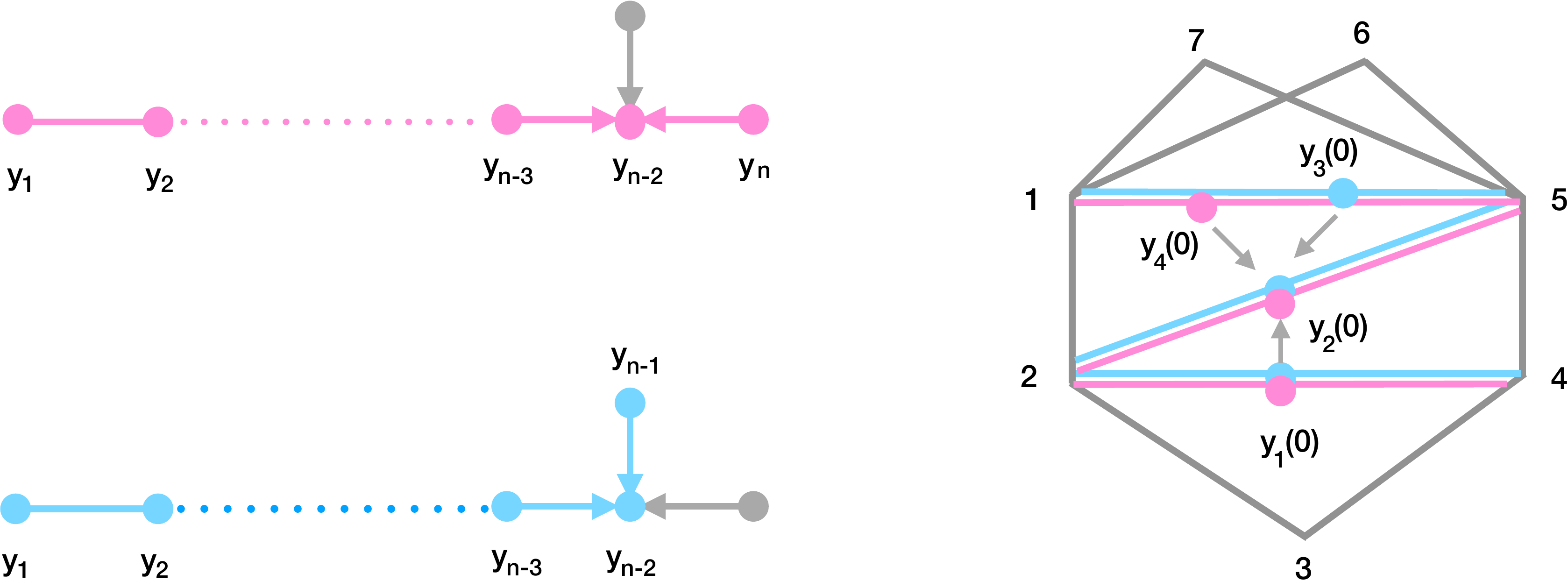}
\caption{(1) The $D_n$ Dynkin diagram is a union of two $A_{n-1}$ sub-diagrams.
(2) The overlapping-poly\-gon representation of the $D_4$ worldsheet. Each $A_{n-1}$ worldsheet corresponds to an $(n+2)$-gon. Each diagonal in the zigzag triangulation corresponds to a node in the Dynkin diagram.}
\label{Dnpolygon}
\end{figure}

The $u$-variable parameterization of the cluster configuration space has the advantage that it provides an invariant description of the moduli space and a binary approach to the boundaries. $Y$-systems further allow us to generalize to cluster configuration spaces of Dynkin type. Nevertheless, it is still desirable to have a description of the moduli space as a configuration space of points. In the following, we discuss a novel construction based on gluing a pair of $A$-type worldsheets.

First, observe that the $D_n$ Dynkin diagram contains two $A_{n-1}$ sub-diagrams that share $n-2$ nodes, as shown on the left of Figure \ref{Dnpolygon}.
We prepare two copies of $(n+2)$-gons and ``glue'' $n+1$ of the common vertices together, leaving the last vertex on each polygon alone, as shown on the right of Figure \ref{Dnpolygon}. The vertices of the two polygons are $(1,2,\dots, n+1,n+2)$ and $(1,2,\dots, n+1,n+3)$, respectively. The positions of the vertices $z_1, \dots, z_{n+3}$ will be our worldsheet variables. We will also denote the unglued vertices $z_{n+2}$ and $z_{n+3}$ as $z_-$ and $z_+$, respectively. We end up with a pair of ``antennae'' corresponding to $z_-$ and $z_+$.
We take zigzag triangulations of the $(n+2)$-gons and associate $y$-variables to the diagonals. The common diagonals $y_1, y_2, \dots, y_{n-2}$ will be identified by the gluing. Note that the $y_{n-1}$ and $y_{n}$ variables are not identified because they belong to the $(1,2,n+1,n+2)$ and $(1,2,n+1,n+3)$ quadrilaterals, respectively. The dihedral coordinates on the $A_{n-1}$ worldsheets define the initial set of $y$-variables that are cross ratios of their respective $z$-variables. By evolving the system according to the $Y$-system equations, we generate all the other variables. We shall see that in addition to the $z_{i}-z_{j}$ factors, new nonlinear factors appear in the cross ratio. The $u$-variables will be expressed as generalized cross ratios of the worldsheet variables.

\begin{figure}[t]\centering
\includegraphics[scale=0.28]{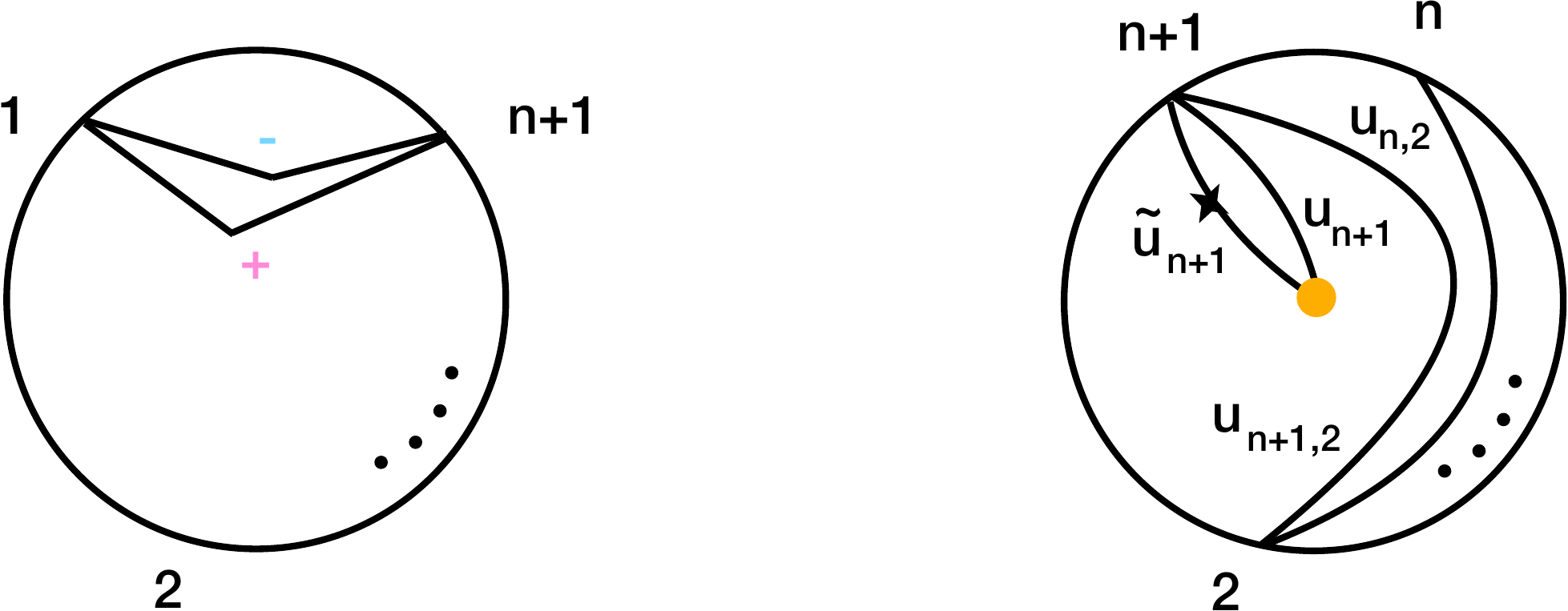}
\caption{One may relate the overlapping-polygon representation of the $D_n$ worldsheet to the once-punctured disc by folding the ``antennae'' inward and identifying $z_{n+1}$ with $z_{1}$ . The extra points $z_{\pm}$ may then be identified with the interior puncture and one uses tagged arcs to distinguish whether the diagonal belongs to the polygon containing $z_+$ or $z_-$.}
\label{Dnconfig}\vspace{-2mm}
\end{figure}

We comment on how this construction may be related to the previous picture of the $D_n$ moduli space as a once-punctured disc~\cite{Fomin_2008}. We draw the first $n+1$ points on the boundary of a disc, as shown on the left of Figure \ref{Dnconfig}. We fold the ``antennae'' inward and then identify~$z_{n+1}$ with $z_1$, which can be thought of as a one-point compactification of the moduli space. The diagonals connecting $z_1$ and $z_{n+1}$ are also folded. We will replace $z_-$ and $z_+$ with a single point and use tagged arcs to distinguish the two diagonals. The plain arc corresponds to the diagonal in the $z_+$ polygon and the notched arc corresponds to the diagonal in the $z_-$ polygon, as shown on the right of Figure \ref{Dnconfig}. Thus we have recovered the construction of the $D_n$ surface cluster algebra as tagged triangulations of the once-punctured disc.

{\bf The $\boldsymbol{D_4}$ example.}
We take the initial $y$- (or $u$-) variables according to the triangulation of Figure \ref{Dnpolygon}:
\begin{align}
\{ u_{1}(0), u_{2}(0), u_{3}(0), u_{4}(0) \}
={}& \left\{\frac{z_{1,4} z_{2,3}}{z_{1,3} z_{2,4}},\frac{z_{1,5} z_{2,4}}{z_{1,4} z_{2,5}}, \frac{z_{4,1} z_{5,6}}{z_{4,6} z_{5,1}},\frac{z_{4,1} z_{5, 7}}{z_{4,7} z_{5,1}}\right\} \nonumber\\
:={}& \{u_{4,2}, u_{5,2}, \widetilde u_{ 5}, u_{5}\}. \label{initial}
\end{align}
In the second line we have labeled the initial $u$-variables according to~\eqref{main}.
We evolve the system according to the $Y$-system equations (\ref{Y-system}) and more specifically~\eqref{localD}. The diagonals are mapped as
\[
u_{i,j} \to u_{i+1,j+1}, \qquad u_i \to \widetilde u_{{i+1}}, \qquad \widetilde u_{ i} \to u_{i+1}.
\]
One important observation is that the indices of $u_{i,j}$, $u_{i}$, $\widetilde u_{ i}$ range from $2 \le i, j \le n+1$, which makes manifest the periodicity of the $D_n$-type $Y$-system. For example, the variable that comes after $u_{5,2}$ would be $u_{2,3}$. At $t=1$, we find
\[
\{u_{5,3}, u_{2,3}, u_{2}, \widetilde u_{ 2}\}=\left\{\frac{z_{2,5} z_{3,4}}{z_{2,4} z_{3,5}}, -\frac{z_{1,5} z_{3,5} z_{2,6} z_{2,7}}{w_{2,3} z_{2,5}}, \frac{z_{1,6} z_{2,5}}{z_{1,5} z_{2,6}}, \frac{z_{1,7}z_{2,5}}{z_{1,5}z_{2,7}} \right\} .
\]
Here a nonlinear factor appears in the expression for $u_{2,3}$ which is a cubic polynomial in~$z_i$'s
\[
w_{i,j} = z_{1,n+3} z_{i,j} z_{n+1,n+2} - z_{1,n+1} z_{i,n+3} z_{j,n+2}.
\]
Noting that $w_{i,j}$ factorizes into a product of $z_{i,j}$'s when $i=1$ or $i=j$, one can conveniently rewrite $u_{2,3}$ as a cross ratio of $w_{i,j}$'s
\[
u_{2,3} = \frac{w_{1,3} w_{2,2}}{w_{1,2} w_{2,3}} .
\]
In this way we generate all the $u$-variables explicitly, noting that they can be elegantly written as cross ratios of $z_{i,j}$'s and $w_{i,j}$'s
\begin{gather*}
\{u_{2,4}, u_{3,4}, \widetilde u_{3}, u_{3}\} =\left\{ \frac{w_{1,4} w_{2,3}}{w_{1,3} w_{2,4}},\frac{w_{2,4} w_{3,3}}{w_{2,3} w_{3,4}},\frac{w_{2,3} z_{3,6}}{w_{3,3} z_{2,6}},\frac{w_{2,3} z_{3,7}}{w_{3,3} z_{2,7}}\right\} ,\\
\{u_{3,5}, u_{4,5}, u_{4},\widetilde u_{ 4}\}
=
\left\{\frac{w_{2,5} w_{3,4}}{w_{2,4} w_{3,5}},\frac{w_{3,5} w_{4,4}}{w_{3,4} w_{4,5}},\frac{w_{3,4} z_{4,7}}{w_{4,4} z_{3,7}},\frac{w_{3,4} z_{4,6}}{w_{4,4} z_{3,6}}
 \right\} ,\\
\{u_{4,2}, u_{5,2}, \widetilde u_{ 5}, u_{5}\}
= \left\{\frac{z_{1,4} z_{2,3}}{z_{1,3} z_{2,4}},\frac{z_{1,5} z_{2,4}}{z_{1,4} z_{2,5}}, \frac{z_{4,1} z_{5,6}}{z_{4,6} z_{5,1}},\frac{z_{4,1} z_{5, 7}}{z_{4,7} z_{5,1}}\right\} .
\end{gather*}
In the final step, we recover the initial set of $u$-variables as guaranteed by periodicity.

Inspired by the result for $D_4$, one can conjecture the form for general $n$.
The $u$-variables are written as cross ratios involving the $z$'s and $w$'s, which can be compactly expressed as~\eqref{main}. It is straightforward to verify that~\eqref{main} is a solution to the local $u$-equations~\eqref{localD}.

\subsection[The positive D\_n moduli space and its boundaries]{The positive $\boldsymbol{D_n}$ moduli space and its boundaries}

The positive region corresponds to a further restriction on the real line and a choice of ordering. We shall denote $z_{n+2} = z_-$ and $z_{n+3} = z_+$ for the two special points. Because the set of worldsheet variables does not include the $z_- - z_+$ factor, $z_{\pm}$ are not ordered with respect to each other. A natural choice of ordering would be $z_{-}, z_{+} \le z_{1} \le \cdots \le z_{n+1}$.
The $y$-variables are positive and the $u$-variables lie in the unit interval. We may also use other orderings, as long as the $u$-variables are positive and lie in the unit interval.

\begin{figure}[t]\centering
\includegraphics[scale=0.35]{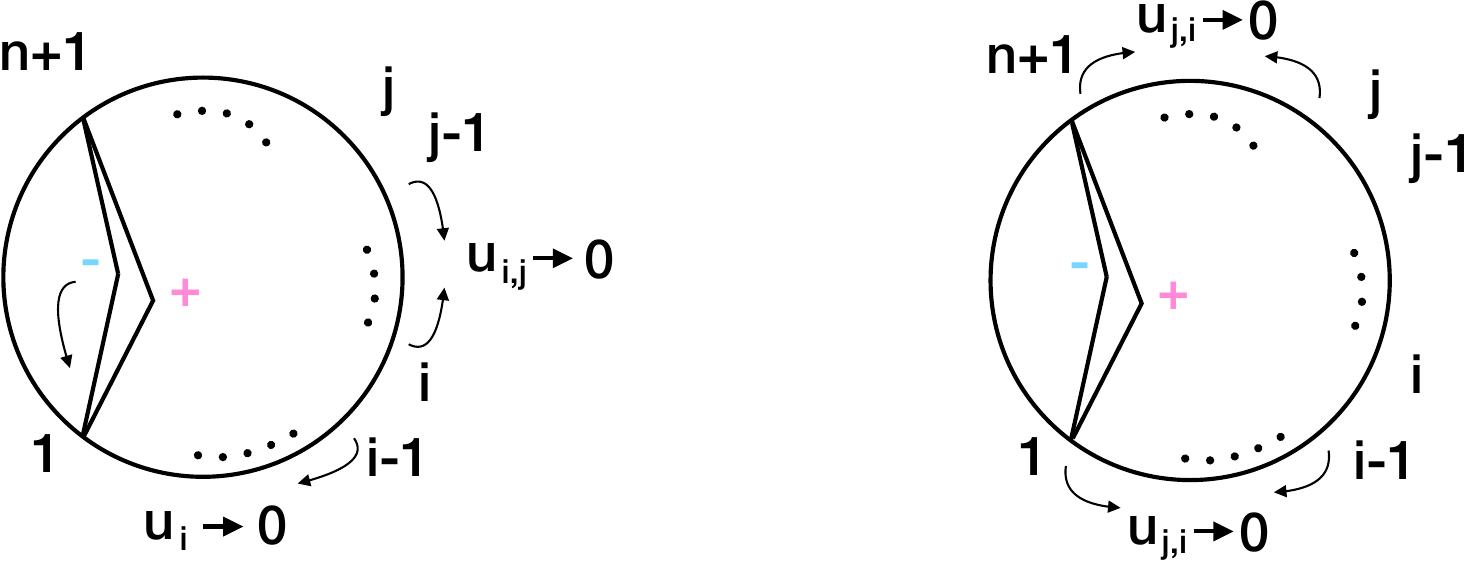}
\caption{The boundary of the $D_n$ moduli space as seen by the worldsheet variables. Each boundary of the positive region corresponds to the pinching of a pair of points and all the points lying in between.}\label{Dnboundary}
\end{figure}

{\bf The boundaries.}
We find the following patterns of pinching take us to the boundaries of the positive moduli space ($1 < j < i \le n+1 $). This is shown in Figure \ref{Dnboundary}.
\begin{itemize}\itemsep=0pt\samepage
\item The $u_{i,j}$ boundary: $z_{i-1}$ pinches with $z_{j}$.
\item The $u_{j,i}$ boundary: $z_{i}$ pinches with $z_{n+1}$, $z_{j-1}$ pinches with $z_1$.
\item The $u_{i}$ boundary: $z_{i-1}$ pinches with $z_{-}$ (but not with $z_{+}$).
\item The $\widetilde u_{ i}$ boundary: $z_{i-1}$ pinches with $z_{+}$ (but not with $z_{-}$).
\end{itemize}
By ``pinching", we mean that the two points, along with all the ordered points in between, are identified.

In the $A_{n}$ configuration space, the pinching of two points always corresponds to a boundary of the moduli space. In the $D_{n}$ configuration space, the pinching of $z_-$ with $z_+$ is not a boundary. Instead, we find the $B_{n-1}$ configuration space as a subspace of $D_n$.

\subsection{The configuration space for other types}

In an upcoming paper, we show that the construction can be extended to the $E_n$ types by gluing a pair of $A_{n-1}$ and $A_{n-2}$ worldsheets. By defining the initial variables using a triangulation and evolving the $Y$-system equations, we can generate solutions in terms of the worldsheet coordinates. For the nonlinear factors of $E_6$, we find sextic polynomials of the form
\begin{gather}
w_{i,j,k,l} = z_{1,9} z_{1,6} z_{i,8} z_{j,7} z_{k,9} z_{l,6}-z_{1,9} z_{1,6} z_{6,9} z_{i,8} z_{j,k} z_{l,7} +z_{1,i} z_{1,6} z_{6,9} z_{j,9} z_{k,8} z_{l,7}\nonumber\\
\hphantom{w_{i,j,k,l} =}{} +z_{1,8} z_{1,9} z_{6,7} z_{6,9} z_{i,l} z_{j,k}.
\label{E6w}
\end{gather}
The $u$-variables can be written as generalized cross ratios of $z_{i,j}$ and $w_{i,j,k,l}$.\footnote{The dihedral coordinates for $E_6$ have also been constructed in connection with the Grassmannian cluster algebra of ${\rm Gr}(4,7)$~\cite{Drummond:2018dfd}, which are related by a sequence of mutations. We directly construct them from the $Y$-systems associated with the $E_6$ Dynkin diagram. It would be interesting to find a map between the worldsheet variables and the cluster $\mathcal{A}$-coordinates.}

By generalizing the cross-ratio parameterization of the $A_{n}$-type $Y$-systems, we may obtain worldsheet parameterizations for all ADE-type $Y$-systems. The non-simply-laced types may be obtained from the simply-laced types by a folding procedure: identifying diagonals in worldsheets in a way consistent with folding the Dynkin diagrams. Further details will be presented in~\cite{WangZhao}.

The gluing method can be formally extended to infinite types whose Dynkin types can be written as a union of $A$-types. For example, the affine $D_n$ diagram can be obtained from two~$A_{n-1}$ diagrams by gluing all but a pair of nodes at each end. This results in overlapping polygons with antennae at both ends. Upon identifying the unglued vertices with internal punctures as described earlier, we recover the affine $D_n$ moduli space as a disc with two internal punctures and $n$ boundary points~\cite{Fomin_2008}. The main difficulty with the infinite types is that the $Y$-system equations generate an infinite number of complicated expressions. It would be an interesting question to understand the form of the polynomials that appear in the $u$-variables and the alphabet.

\subsection{Summary of alphabets}
Once we obtain the worldsheet description of $u$-variables in terms of polynomial factors of $z$-variables, we find the alphabets by fixing the ${\rm SL}(2)$ redundancy. We summarize the results for the alphabets of finite types below.

The $A_{n}$ alphabet may be obtained by fixing three points in the $z_i - z_j$ factors as $z_1 = -1$, $z_2 = 0$, and $z_{n+3} = \infty$:
\[
\Phi_{A_n}(z_3, z_4, \dots, z_{n+2}) = \bigcup_{3 \le i \le n+2} \{z_i, 1+z_i \} \cup \bigcup_{3 \le i < j \le n+2} \{z_{i,j}\} ,
\]
Using the same canonical gauge choice ($z_1 = -1$, $z_2 = 0$, $z_{n+1} = \infty$), we recover the $D_{n}$ alphabet found in~\cite{Chicherin:2020umh}
\begin{gather}
\Phi_{D_n}(z_3, z_4, \dots, z_{n}, z_{-}, z_{+}) = \Phi_{A_{n-1}}(z_3, z_4, \dots, z_{n}, z_{-}) \cup \{z_+, 1+z_+\} \nonumber\\
\hphantom{\Phi_{D_n}(z_3, z_4, \dots, z_{n}, z_{-}, z_{+}) =}{}\cup \bigcup_{3 \le i \le n} \{z_{i,+}, z_i + z_{-}z_{+} \}\nonumber\\
\hphantom{\Phi_{D_n}(z_3, z_4, \dots, z_{n}, z_{-}, z_{+}) =}{}\cup \bigcup_{3 \le i < j \le n} \{-z_i+z_j+z_i z_j-z_i z_{-}-z_i z_{+}+z_{-} z_{+} \} .
\label{Dnalphabet}
\end{gather}
An alphabet for $B_{n-1}$ follows from $D_n$ by a simple folding identifying $z_- = z_+$. An alphabet for~$C_{n-1}$ has appeared explicitly in~\cite{Chicherin:2020umh}. We find that it can be obtained from the $A_{2n-2}$ alphabet by the identification $z_{2n+1-i} = z_{n+1}/z_{n+1-i}$ for $i = 1,\dots, n-2$.
They are polynomials that are at most quadratic in the $z$-variables
\begin{gather}
\Phi_{B_{n-1}}(z_3, z_4, \dots, z_{n}, z_{-}) = \Phi_{A_{n-1}}(z_3, z_4, \dots, z_{n}, z_{-}) \cup \bigcup_{3 \le i \le n} \big\{z_i + z_{-}^2 \big\} \nonumber\\
\hphantom{\Phi_{B_{n-1}}(z_3, z_4, \dots, z_{n}, z_{-}) =}{}\cup \bigcup_{3 \le i < j \le n} \bigl\{-z_i+z_j + z_i z_j-2z_i z_{-}+z_{-}^2 \bigr\} , \nonumber\\
\Phi_{C_{n-1}}(z_3, z_4, \dots, z_{n}, z_{-}) =\Phi_{A_{n-1}}(z_3, z_4, \dots, z_{n}, z_{-})\cup \bigcup_{3\le i \le j \le n} \{ z_i z_j + z_- \}
.
\label{BCalphabet}
\end{gather}

Alphabets for $E_6$, $F_4$ and $G_2$ are not known in the literature. A detailed derivation will be given in~\cite{WangZhao} and here we summarize the results. An $E_6$ alphabet is obtained from the sextic polynomial (\ref{E6w}) by fixing three points. In the canonical gauge choice $(z_1 = -1, z_2 = 0, z_{6} = \infty)$, it consists of 42 letters that are polynomials with degree at most 4:
\begin{gather}
\Phi_{E_6}(z_3, z_4, z_5, z_7, z_8, z_9) = \Phi_{A_5} (z_3, z_4, z_5, z_7, z_8) \cup \{z_9, 1+z_9\} \} \nonumber\\
\hphantom{\Phi_{E_6}(z_3, z_4, z_5, z_7, z_8, z_9) =}{}\cup \bigcup_{3\le i \le 5} \{z_{i,9}, z_{i}+z_7 z_9, z_{i}+z_8 z_9 \nonumber\\
\hphantom{\Phi_{E_6}(z_3, z_4, z_5, z_7, z_8, z_9) =}{}\cup \bigcup_{3\le i < j \le 5} \{-z_i+z_j+z_i z_j- z_i z_7- z_i z_9 + z_7 z_9, -z_i+z_j+z_i z_j \nonumber\\
\hphantom{\Phi_{E_6}(z_3, z_4, z_5, z_7, z_8, z_9) =\cup \bigcup_{3\le i < j \le 5}\{}{} - z_i z_8- z_i z_9 + z_8 z_9, \nonumber\\
\hphantom{\Phi_{E_6}(z_3, z_4, z_5, z_7, z_8, z_9) =\cup \bigcup_{3\le i < j \le 5}\{}{} z_i z_j-z_i z_7+z_i z_8-z_j z_8 + z_i z_8 z_9-z_7 z_8 z_9 \} \nonumber\\
\hphantom{\Phi_{E_6}(z_3, z_4, z_5, z_7, z_8, z_9) =}{}\cup \big\{-z_3 z_4+z_3 z_7+z_4 z_5-z_4 z_7+z_4 z_8-z_5 z_8+z_3 z_4 z_5-z_3 z_4 z_7 \nonumber\\
\hphantom{\Phi_{E_6}(z_3, z_4, z_5, z_7, z_8, z_9) =\cup \{}{} -z_3 z_4 z_9-z_3 z_5 z_8+z_3 z_7 z_8+z_3 z_7 z_9+z_4 z_8 z_9-z_7 z_8 z_9, \nonumber\\
\hphantom{\Phi_{E_6}(z_3, z_4, z_5, z_7, z_8, z_9) =\cup \{} -z_3 z_5+z_4 z_5+z_3 z_4 z_5-z_3 z_4 z_7+z_3 z_4 z_8-z_3 z_5 z_8-z_3 z_5 z_9 \nonumber\\
\hphantom{\Phi_{E_6}(z_3, z_4, z_5, z_7, z_8, z_9) =\cup \{}-z_3 z_8 z_9+z_4 z_7 z_9+z_5 z_8 z_9 +z_3 z_4 z_8 z_9-z_3 z_7 z_8 z_9-z_3 z_8 z_9^2 \nonumber\\
\hphantom{\Phi_{E_6}(z_3, z_4, z_5, z_7, z_8, z_9) =\cup \{} +z_7 z_8 z_9^2
\big\}
.
\label{E6alphabet}
\end{gather}
An $F_4$ alphabet may be obtained by ``folding'' the $E_6$ alphabet~\eqref{E6alphabet} as ($z_7 = -{z_5}/{z_3}$, $z_8 = -{z_5}/{z_4}$) and examining all the polynomial factors that appear. It consists of 28 polynomial letters with degree at most 5:
\begin{gather*}
\Phi_{F_4} (z_3, z_4, z_5, z_9)= \Phi_{A_4}(z_3, z_4, z_5, z_9) \cup \bigcup_{3\le i \le j \le 4} \{z_i z_j + z_5, z_i z_j - z_5 z_9\} \\
\cup \big\{-z_3^2+z_3 z_4+z_3 z_5-z_5 z_9 +z_3^2 z_4-z_3^2 z_9, -z_3^2+2 z_3 z_5-z_5 z_9 + z_3^2 z_5-z_3^2 z_9 ,
\\
-z_3 z_4+z_3 z_5+z_4^2-z_5 z_9 + z_3 z_4^2-z_3 z_4 z_9, -z_3 z_4+z_3 z_5+z_4 z_5-z_5 z_9 + z_3 z_4 z_5-z_3 z_4 z_9
\\
-z_4^2+2 z_4 z_5 -z_5 z_9 + z_4^2 z_5-z_4^2 z_9, -z_3^2 z_5+2 z_3 z_4 z_5-z_5^2 z_9 + z_3^2 z_4^2-z_3^2 z_5 z_9, \\
-2 z_3^2 z_4 +2 z_3 z_4^2 + z_3^2 z_5 + z_3^2 z_9 -2 z_3 z_5 z_9 -z_4^2 z_9 +z_5 z_9^2+ z_3^2 z_4^2-2 z_3^2 z_4 z_9+z_3^2 z_9^2 , \\
-2 z_3 z_4 z_5+2 z_3 z_5^2+z_4^2 z_5-z_5^2 z_9 -z_3^2 z_4^2+2 z_3 z_4^2 z_5+z_3^2 z_5^2-2 z_3 z_4 z_5 z_9 + z_3^2 z_4^2 z_5-z_3^2 z_4^2 z_9
 \big\}
 .\end{gather*}
A $G_2$ alphabet may be obtained by folding the $B_3$ alphabet~\eqref{BCalphabet} as $z_- = -z_4/z_3$ and consists of 8 polynomial letters with degree at most 4:
\begin{gather}
\Phi_{G_2}(z_3, z_4) = \big\{z_3, z_4, 1+z_3, 1+z_4, z_3-z_4, z_3^2\!+\!z_4, z_3^3\!+\!z_4^2, -z_3^3\!+\!z_4^2\!+\!3 z_3^2 z_4\!+\! z_3^3 z_4\big\}.\!\!
\label{G2}
\end{gather}

\section[Cluster string integrals and scattering equations in z-variables]{Cluster string integrals and scattering equations in $\boldsymbol{z}$-variables}\label{sec3}
Having discussed the configuration spaces for finite-type cluster algebras, now we move to the study of cluster string integrals and the associated saddle-point equations in $z$-variables. We will focus on the type-$D$ case but the discussion directly extends to any other finite type.

\subsection{Cluster string integrals and their factorization}
From the general form of cluster (open-) string integrals, all we need to do is to express the ``Koba--Nielsen'' factor, the canonical forms and the integration domain in worldsheet-like variables; given the map from $u$-variables as discussed above, all of these can be obtained by translating those in $u$-variables. To keep the notation compact, we set $\alpha'=1$ in the following. The Koba--Nielsen factor for the $D_n$ type takes the form
\begin{gather}
\prod_{I} u_{I}^{X_{I}} = \prod_{1\le j < i \le n+1} z_{i,j}^{-{c_{i, j}}} \prod_{1< j < i < n+1} w_{j,i}^{-c_{j,i}} \prod _{i=1}^{n+1} z_{i,+}{}^{-c_i} z_{i,-}{}^{-\widetilde c_{{i}}}
\label{KNDn}
.
\end{gather}
Recall that there is an $X_I$ kinematical variable for each $u_I$ variable. All the $c_{i,j}$, $c_i$, $\widetilde c_{ i}$ variables are linear combinations of the $X_I$ variables{\samepage
\begin{gather}
c_{i, j} = X_{i, j} + X_{i+1,j+1} - X_{i, j+1} - X_{i+1, j} \quad \text{for non-adjacent }1\le i,j \le n+1 , \nonumber\\
c_{n+1,1} = \sum_{i=1}^n \big(X_i + \widetilde X_{ i} - X_{i,i+1} \big) , \nonumber\\
c_{i, i+1} = X_{i, i+1} + X_{i+1,i+2} - X_{i, i+2} - X_{i+1} - \widetilde X_{{i+1}} \quad \text{for } 1\le i \le n , \nonumber\\
c_{i} = X_i + \widetilde X_{{i+1}} - X_{i,i+1}, \quad \widetilde c_{ i} = \widetilde X_{ i} + X_{i+1} - X_{i,i+1} \quad \text{for } 1 < i < n+1,\nonumber\\
c_{1} = -X_{2}, \qquad \widetilde c_{ 1} = - \widetilde X_{ 2}, \qquad c_{n+1} = -\widetilde X_{ 1}, \qquad \widetilde c_{{n+1}} = - X_1 .
\label{ABHYDn}
\end{gather}
Here $X_i$ are defined cyclically, e.g., $X_{n+1} := X_{1}$, and $X_{i,i} = X_{i,i-1} = 0$.}

As mentioned, the most basic property of such stringy integrals is that the convergence domain is given by a generalized associahedron. In the $A_{n-3}$ case, the variables $X_{i,j} =(p_i + p_{i+1}+ \cdots + p_{j-1})^2$ are called planar variables, and $c_{a,b} = -(p_a + p_b)^2$ are the Mandelstam variables (with a minus sign). By requiring all $X_{i,j} \geq 0$ and asking all but $n{-}3$ $c_{a,b}$ to be positive constants, they carve out an ABHY subspace that is a ``kinematic associahedron'' of type $A_{n-3}$ which gives the convergence domain, and its boundaries exactly encode the factorization properties of the stringy integral~\cite{Arkani-Hamed:2017mur}.\footnote{Note that there are same number of $c_{a,b}$ and $X_{i,j}$, thus one must pick $n{-}3$ $c_{a,b}$ not to be set to constants but rather as basis variables for the associahedron; see~\cite{Arkani-Hamed:2017mur, Arkani-Hamed:2019vag} for precisely which $n{-}3$ $c_{a,b}$ cannot be chosen as constants, which are dictated by the ``mesh'' or spacetime picture.} The $D_n$ case is completely analogous: by setting all but $n$ of $c_I$ as positive constants and requiring all $X_I \geq 0$, they carve out a generalized associahedron of type $D_n$, which gives the convergence domain, and its boundaries are in one-to-one correspondence with the boundaries of the configuration space that we studied in the previous section~\cite{Arkani-Hamed:2017mur, Arkani-Hamed:2019vag}. In a close connection with $Y$-systems, the equations can also be regarded as an evolution in a discrete spacetime with extra boundary conditions. We will come back to such ABHY generalized associahedron as we discuss the pushforward map from considering saddle-point equations.

Note that when we study stringy integrals, it is always convenient to fix the gauge, e.g., in the Koba--Nielsen factor exponents of $z_{n+1}$ sum to zero and decouple. Next, we need to find the positive part ${\mathcal M}^+_{D_n}$ and its canonical forms, when all the $u$-variables are positive (thus between~$0$ and $1$). Since all $Y=u/(1-u)$ are manifestly positive given positive initial $Y$'s, e.g.,~\eqref{initial}, it is easy to see that this is equivalent to the following ordering: $z_1< z_2< \cdots <z_{n{+}1}$ and $z_\pm <z_1$ (note that there is no ordering between $z_+$ and $z_-$). The canonical form can be obtained as the wedge product of ${\rm d}\log y$ for any initial cluster with {\it acyclic} quiver, e.g., those in~\eqref{initial}. It is a~remarkable fact that the result does not depend on which initial cluster we start with, and for convenience we work with gauge-fixed variables, e.g., with $z_{n{+}1} \to \infty$, and we obtain a compact ``Parke--Taylor-like'' form in the remaining $z$-variables
\[
\omega_{D_n}:=\frac{\prod_{i=\pm, 3}^n {\rm d} z_i}{z_{1,+} z_{1,-} z_{2,3} z_{3,4} \cdots z_{n{-}1, n}} ,
\]
where in our gauge-fixing $z_{1,\pm}=1+ z_\pm$ and $z_{2,3}=z_3$ (also $z_{1,2}=1$). Thus we have written all the ingredients of the $D_n$ string integral in $z$-variables.

Returning to the $D_4$ example, one may readily write down the stringy integral~\eqref{stringyintegral} in terms of the worldsheet variables
\begin{gather*}
I_{D_4} =
\int_{0 < z_3 < z_4 \atop z_\pm < -1} \frac{{\rm d}z_3 {\rm d}z_4 {\rm d}z_{-} {\rm d}z_{+}}{(1+z_-)(1+z_+)z_3 z_{3,4}} z_3^{X_{4,2}} z_4^{X_{1,2}-X_{1,3}-X_{4,2}} z_-^{-\widetilde X_{{2}}-X_3+X_{2,3}} z_{+}^{-\widetilde X_{{2}}-X_3+X_{2,3}} \\
\hphantom{I_{D_4} =}{}\times (1+z_3 ){}^{-X_{3,1}+X_{4,1}-X_{4,2}} (1+z_4 ){}^{X_1+\widetilde X_{{1}}-X_{1,2}-X_{4,1}+X_{4,2}} (1+z_{-} ){}^{\widetilde X_{ 2}} (1+z_{+} ){}^{X_{2}} \\
\hphantom{I_{D_4} =}{}\times z_{3,4}{}^{X_{1,3}} z_{3,-}{}^{-\widetilde X_{{3}}+X_{3,4}-X_4} z_{3,+}{}^{-X_3-\widetilde X_{{4}}+X_{3,4}}z_{4,-}{}^{-X_1-\widetilde X_{{4}}+X_{4,1}} z_{4,+}{}^{-\widetilde X_{{1}}-X_4+X_{4,1}} \\
\hphantom{I_{D_4} =}{}\times (z_3+z_{-} z_{+} ){}^{X_3+\widetilde X_{{3}}-X_{2,3}+X_{2,4}-X_{3,4}} (z_4+z_{-} z_{+} ){}^{-X_{2,4}-X_{3,1}+X_{3,4}}\\
\hphantom{I_{D_4} =}{}\times (z_3-z_4-z_3z_4 +z_3z_{-} +z_3z_{+} -z_{-} z_{+} ){}^{X_4+\widetilde X_{{4}}+X_{3,1}-X_{3,4}-X_{4,1}} .
\end{gather*}
We may see factorization into lower-point amplitudes at the boundaries of the moduli space. The $u_{1,3}$ boundary corresponds to $z_4 \to z_3$, where $X_{1,3}$ must also tend to 0 and the stringy integral reduces to the $D_3$ integral
\begin{gather*}
\int_{0 < z_3 \atop z_\pm < -1} \frac{{\rm d}z_3 {\rm d}z_{-} {\rm d}z_{+}}{z_3(1+z_-)(1+z_+)} z_3^{X_{1,2}-X_{1,3}} z_-^{-\widetilde X_{{2}}-X_3+X_{2,3}}z_+^{-X_2-\widetilde X_{{3}}+X_{2,3}}
\left(1+z_3\right){}^{X_1+\widetilde X_{{1}}-X_{1,2}-X_{3,1}} \\
\quad \times \left(1+z_-\right){}^{\widetilde X_{{2}}}\left(1+z_{+}\right){}^{X_{2}} z_{3,-}{}^{-X_1-\widetilde X_{{3}}+X_{3,1}} z_{3,+}{}^{-\widetilde X_{{1}}-X_3+X_{3,1}}\left(z_3+z_{-} z_{+}\right){}^{X_3+\widetilde X_{{3}}-X_{2,3}-X_{3,1}} .
\end{gather*}
Similarly for the $z_3 \to 0$. At the $z_+ \to -1$ (or $z_- \to -1$) poles, the $D_4$ integral reduces to the~$A_3$ integral, which is equivalent to the $D_3$ integral but appears in a different parameterization
\begin{gather*}
\int_{0 < z_3 < z_4 \atop z_- < -1}
\frac{{\rm d}z_3 {\rm d}z_4 {\rm d}z_{-}}{z_3 z_4 (1+z_-)} z_3^{X_{4,2}} z_4^{X_{1,2}-X_{1,3}-X_{4,2}} z_-^{-\widetilde X_{{2}}+X_{2,3}-X_3} \left(1+z_3\right){}^{-X_{4,2}-X_3+X_4} \\
\qquad \times \left(1+z_4\right){}^{-X_{1,2}+X_{4,2}+X_1-X_4}\left(1+z_-\right){}^{\widetilde X_{{2}}} z_{3,4}{}^{X_{1,3}} z_{3,-}{}^{-X_{2,3}+X_{2,4}+X_3-X_4} z_{4,-}{}^{-X_{2,4}-X_1+X_4} .
\end{gather*}

For other types, the stringy integral can be obtain in an analogous manner. The Parke--Taylor form is the wedge product of ${\rm d}\log$ of all the initial $y$-variables, written in terms of the alphabets. The Koba--Nielsen factor can be obtained by expressing the $u$-variables as cross ratios of the alphabets. This determines the linear combinations of the kinematical variables $X_I$'s that appear in the exponents of the alphabets. The $B_3$ stringy integral is
\begin{gather*}
I_{B_3} =
\int_{0 < z_3 < z_4 \atop z_- < -1} \frac{{\rm d}z_3 {\rm d}z_4 {\rm d}z_{-}}{(1+z_-) z_3 z_{3,4}} z_3^{X_{4,2}} z_4^{X_{1,2}-X_{1,3}-X_{4,2}} z_-^{2 X_{2,3}-X_2-X_3} \\
\hphantom{I_{B_3} =}{} \times (1+z_3 ){}^{-X_{3,1}+X_{4,1}-X_{4,2}} (1+z_4 ){}^{-X_{1,2}-X_{4,1}+X_{4,2}+X_1} (1+z_- ){}^{X_2} \\
\hphantom{I_{B_3} =}{} \times z_{3,4}{}^{X_{1,3}} z_{3,-}{}^{2 X_{3,4}-X_3-X_4} z_{4,-}{}^{2 X_{4,1}-X_1-X_4} \big(z_3 + z_-^2 \big){}^{-X_{2,3}+X_{2,4}-X_{3,4}+X_3} \\
\hphantom{I_{B_3} =}{} \times\big(z_4 + z_-^2 \big){}^{-X_{2,4}-X_{3,1}+X_{3,4}} \bigl(-z_3+z_4 +z_3 z_4 -2 z_3 z_- + z_-^2 \bigr){}^{X_{3,1}-X_{3,4}-X_{4,1}+X_4}
 .
\end{gather*}
At the same $u_{1,3}$ boundary when $z_4 \to z_3$, it reduces to the $B_2$ integral
\begin{gather*}
\int_{0 < z_3 \atop z_- < -1} \frac{{\rm d}z_3 {\rm d}z_{-}}{(1+z_-) z_3}
z_3^{X_{1,2}-X_{1,3}} z_-^{2 X_{2,3}-X_2-X_3} (1+z_- ){}^{X_2} (1+z_3 ){}^{-X_{1,2}-X_{3,1}+X_1} \\
\qquad \times(z_- - z_3 ){}^{2 X_{3,1}-X_1-X_3} \big(z_-^2+z_3 \big){}^{-X_{2,3}-X_{3,1}+X_3} .
\end{gather*}
At the $z_- \to -1$ pole, the $B_3$ integral reduces to the $A_2$ integral
\begin{gather*}
\int_{0 < z_3 < z_4} \frac{{\rm d}z_3 {\rm d}z_4}{z_3 z_{3,4}}
z_3^{X_{4,2}} z_4^{X_{1,2}-X_{1,3}-X_{4,2}} (1+z_3){}^{-X_{2,3}+X_{2,4}-X_{4,2}}(1+z_4){}^{-X_{1,2}-X_{2,4}+X_{4,2}} z_{3,4}{}^{X_{1,3}} .
\end{gather*}

Finally, the $G_2$ stringy integral is
\begin{gather}
I_{G_2} = \int_{0 < z_3 < z_4} \frac{{\rm d}z_3 {\rm d}z_4}{z_3 z_{3,4}} z_3^{X_{4,2}} z_4^{X_{1,2}-X_{1,3}+2 X_{2,3}-X_{2,4}-X_{3,1}+X_{3,4}+2 X_{4,1}-X_{4,2}} \nonumber\\
\quad\times(1+z_3){}^{-X_{3,1}+3 X_{4,1}-X_{4,2}} (1+z_4){}^{-X_{1,2}-X_{4,1}+X_{4,2}} (z_3-z_4){}^{X_{1,3}} \big(z_3^2+z_4\big){}^{-X_{2,4}-X_{3,1}+3 X_{3,4}} \nonumber\\
\quad\times\big(z_3^3+z_4^2\big){}^{-X_{2,3}+X_{2,4}-X_{3,4}} \bigl(-z_3^3+z_4^2+3 z_3^2 z_4+ z_3^3 z_4\bigr){}^{X_{3,1}-X_{3,4}-X_{4,1}}
\label{G2integral}
.
\end{gather}
One may readily check that it reduces drastically to the $A_1$ integral when $z_4 \to z_3$
\[
\int_{0 < z_3} \frac{{\rm d}z_3}{z_3} z_3^{X_{1,2}-X_{1,3}} (1+z_3){}^{-X_{1,2}-X_{2,3}} .
\]

\subsection{The scattering equations of finite type and numbers of solutions}
The so-called scattering equations arise as the saddle-point equations of the stringy integrals. The $D_n$ scattering equations can be read off from the Koba--Nielsen factor~\eqref{KNDn} as
\[
\sum_ {1\le j < i \le n+1} {c_{i, j}} \, {\rm d}\log z_{i,j} + \sum_ {1< j < i < n+1} c_{j,i} \,{\rm d}\log w_{j,i} + \sum_{1 \le i \le n+1} \big(c_i \,{\rm d}\log z_{i,+} + \widetilde c_i \,{\rm d}\log z_{i,-} \big) = 0.
\]
Note that with gauge-fixing, e.g., $z_1=-1$, $z_2=0$, $z_{n+1} = \infty$, we have only $n$ variables, $z_3, \dots, z_n$ and $z_\pm$, and $n$ equations. An important comment is that by pulling back the scattering equations to any ABHY subspace explained below~\eqref{ABHYDn}, we will have a diffeomorphism from the positive part of the space to the interior of the ABHY generalized associahedron~\cite{Arkani-Hamed:2017mur}. For example, for~$D_4$, there are $4^2=16$ $X$'s. By choosing a subspace constraint, for instance, by choosing exactly the $12$ combinations of~$X$'s given in the first three lines of~\eqref{ABHYDn} as positive constants, only $4$~$X$'s are left as variables and we find a map from $(z_3, z_4, z_+, z_-)$ to $(X_{4,2}, X_{1,3}, X_2, \widetilde X_{{2}})$. It is straightforward to check that this maps the positive part of the space to the interior of the ABHY $D_4$ polytope.

Another important question regards the number of solutions for such scattering equations, or the number of saddle points. Recall that for type $A_{n-3}$, the number of saddle points is $(n-3)!$, which can be nicely interpreted as $(n-3)!$ orderings of $n-3$ ``particles'' on a line, e.g., labeled by $z_2, \dots, z_{n{-}2}$ and bounded by $z_1=0$, $z_{n{-}1}=1$ with a ``potential'' given by the $\log$ of the Koba--Nielsen factor~\cite{Cachazo:2016ror, Kalousios:2013eca}. It is a remarkable fact that given real Mandelstam variables (which are coupling constants in the potential), these $(n{-}3)!$ solutions are all real. Now it is an open question if we can interpret saddle points, or solutions of scattering equations, for other finite-type cluster string integrals in this way. If so, we might find physical applications for them as color factors or open string disk integrals~\cite{Koba:1969kh} for $A_{n-3}$ string integrals. As an illustrative example, the $G_2$ scattering equations can be written explicitly as
\begin{gather*}
\frac{c_1}{z_3}+\frac{c_2}{1+z_3}+\frac{c_5}{z_3-z_4}+\frac{2 c_6 z_3}{z_4+z_3^2}+\frac{3 c_7 z_3^2}{z_4^2+z_3^3}+\frac{3c_8 z_3 \left(-z_3+2 z_4+z_3 z_4\right)}{-z_3^3+z_4^2+3 z_3^2 z_4+ z_3^3 z_4} = 0 , \\
\frac{c_2}{z_4}+\frac{c_4}{1+z_4}-\frac{c_5}{z_3-z_4}+\frac{c_6}{z_4+z_3^2}+\frac{2 c_7 z_4}{z_4^2+z_3^3}+\frac{c_8 \left(2 z_4 +3 z_3^2+ z_3^3\right)}{-z_3^3+z_4^2+3 z_3^2 z_4+ z_3^3 z_4} = 0.
\end{gather*}
Here $c_i$ are the exponents that appear in the stringy integral~\eqref{G2integral}. The system generically has $13$~solutions. As we will show in the next section, out of all $25$ connected components of the configuration space we find exactly $13$ bounded regions. However, if we choose real values of $c_i$, we only find numerically $9$ real solutions for $(z_3, z_4)$ (and two pairs of complex solutions).

The problem of finding the number of saddle points becomes computationally difficult for higher-rank cases. We may numerically search for the solutions using the package HomotopyContinuation.jl~\cite{Breiding_2018} in the software Julia (see some applications in physics in~\cite{Agostini:2021rze,Sturmfels:2020mpv}).\footnote{The main idea of the package HomotopyContinuation.jl is to solve polynomial equations using homotopy continuation. It is easy to construct
certain parameters $c$ such that there is a solution $z$ for a set of equations~$F(c,z)$ with the parameters $c$. If there is a continuous way to transform $c$ to another set of parameters $c'$, the package will track the path and find the new solutions $z'$ for the new set of equations $F(c',z)$. There is an even fancier way to find more solutions called monodromy. The idea is to continue to transform $c'$ to another set of parameters~$c''$, and transform $c''$ back to $c'''=c$. Correspondingly, the package will find solutions $z''$ for $c''$, and $z'''$ for $c'''=c$ while $z'''$ could be a new solution for the original parameters $c$. Repeating such kind of loops for all solutions already obtained until there are no new solutions after 10 loops (by default), the package will gather all solution obtained and treat them as the whole set of the solutions for $F(c,z)$. Using homotopy continuation again, the package can find all solutions $z^*$ for any given parameters $c^*$.
See more in \url{https://www.juliahomotopycontinuation.org/guides/introduction/}.
}
For $D_n$, we found the following numbers of solutions to the scattering equations:
\begin{gather}
\label{numofsol}\renewcommand{\arraystretch}{1.2}
\begin{tabular}{c|c|c|c|c}
 $D_4$ & $D_5$ & $D_6$ & $D_7$ & $D_8$ \\
 \hline
 55 & 674 & 10215 & 183256& $\geq$3787649
\end{tabular}
\end{gather}
These numerical calculations can be completed in several hours for $D_n$ up to $n=7$ while for~$D_8$, we aborted the computation after the software has struggled to search for the remaining possible solutions for several days.
So we can only confirm that there are at least 3787649 solutions for~$D_8$.\footnote{The solutions found in Julia usually only have accuracy at order ${\mathcal O}\big(10^{-16}\big)$ due to the use of floating numbers but it is very convenient to increase their accuracy to order, for example, ${\mathcal O}\big(10^{-100}\big)$ in Mathematica. We have verified the solutions for $D_4$, $D_5$, $D_6$, $D_7$, $F_4$, $E_6$ both in Julia at order ${\mathcal O}\big(10^{-16}\big)$ using its built-in commands and in Mathematica at order ${\mathcal O}\big(10^{-100}\big)$. We have verified 3787649 solutions for $D_8$ in Julia.}
We also found 2411 solutions for $F_4$ and 51466 solutions for $E_6$.

\section{Topological properties of the spaces}\label{sec4}
In this section, we study the topology of cluster configuration spaces using a finite-field method. Recall that this has been initiated in~\cite{Arkani-Hamed:2020tuz} for at least types $A_n$, $B_n$ and $C_n$, where for each case the number of points in the space over $\mathbb{F}_p$ is shown to be a polynomial of $p$. Let us list these polynomials for completeness
\begin{gather*}
|{\mathcal M}_{A_n}({\mathbb F}_p)| = (p-2)(p-3)\cdots (p-n-1) ,
\\
|{\mathcal M}_{B_n}({\mathbb F}_p)| = (p-n-1)^n ,
\\
|{\mathcal M}_{C_n}({\mathbb F}_p)| = (p-n-1) (p-3)(p-5) \cdots (p-2n+1) .
\end{gather*}
Note that when all the letters can be written as linear factors, the space can be realized as a hyperplane arrangement complement and it is known that we have polynomials for point counts. This is the case for type $A$ in terms of worldsheet variables and type $B$ using certain ``linear'' variables as in~\cite{Arkani-Hamed:2020tuz} (for type $C$ we have not been able to determine whether or not it can be realized as a hyperplane arrangement complement). In such hyperplane cases, it is well known that the coefficients of point-count polynomial give dimensions of the $k$-th cohomology for $k=1, 2, \dots, n$, which are exactly the number of independent $k$-forms generated by wedging ${\rm d}\log$ of these letters. By plugging in $p=1$ we find the Euler characteristics for types $ABC$ as expected, and also by plugging in $p=-1$, we obtain the number of connected components.\footnote{Note that the Dynkin classification is with respect to the cluster algebra, not the crystallographic Coxeter arrangement. For example, the braid arrangement corresponding to the Coxeter group of type $A_{n}$ would reduce to our $A_{n-3}$ space after an ${\rm SL}(2)$ gauge-fixing.}

However, for type $D_n$ and exceptional cases, it is expected that the point count will no longer be polynomials (thus the space cannot be realized as hyperplane arrangement complement), and it is not clear how to get dimensions of cohomology, etc. in these cases. In this section, we mainly focus on the type-$D_n$ case and perform numerical ``experiments'' of counting points for a sufficient number of prime numbers, all the way up to $n=10$. We then conjecture that the point count is always a {\it quasi-polynomial} of $p$ with special ``correction terms". Our conjecture thus yields predictions for dimensions of cohomology for the $D_n$ space, which up to $n=6$ are supported by the number of saddle points we find. Note that in general we do not expect the cohomology to be generated by ${\rm d}\log$ forms only, but we conjecture that this is still the case for 2-forms and 3-forms and we propose an all-$n$ formula for these dimensions. Finally, we will summarize all the linear relations we find for ${\rm d}\log$ 2-forms, which are useful for bootstrap based on such an alphabet.

Before proceeding, we also review the simple example of $G_2$, whose point count was found to be a quasi-polynomial using $F$-polynomials in~\cite{Arkani-Hamed:2020tuz}. It is easy to see that the same quasi-polynomial can be found by using the $8$ letters in~\eqref{G2}
\begin{gather*}
|{\mathcal M}_{G_2}({\mathbb F}_p)|= \begin{cases} (p-4)^2 & \mbox{if $p = 2 \mod 3$},\\
(p-4)^2 + 4& \mbox{if $p = 1 \mod 3$}.
\end{cases}
\end{gather*}
Note that by plugging in $p=-1$, the point count predicts that there are $25$ connected components, which agrees with the number of sign patterns of the $8$ letters and can be visualized by directly plotting them; by plugging in $p=1$ it gives the Euler characteristic as $13$, which agrees with the number of saddle points as we found above.

\subsection[Point count of D\_n over F\_p]{Point count of $\boldsymbol{D_n}$ over $\boldsymbol{\mathbb{F}_p}$}

As explained in~\cite{Arkani-Hamed:2020tuz}, the point count shows that ${\mathcal M}_{D_n}$ cannot be a hyperplane arrangement complement already for $n=4,5$. For $D_4$ and $D_5$, by counting the number of points in the space over $\mathbb{F}_p$ for prime numbers $p \ge 5$, it was previously found that they can be interpolated by the quasi-polynomials
\begin{gather*}
\begin{split}
&|{\mathcal M}_{D_4}({\mathbb F}_p)|=p^4-16 p^3+93 p^2-231 p+207 + \delta_3(p) ,
\\
&|{\mathcal M}_{D_5}({\mathbb F}_p)|=p^5-25 p^4+244 p^3-1156 p^2+2649 p-2355 +\delta_3(p) \left(5 p-36\right)-\delta_4(p) .
\end{split}
\end{gather*}
where $\delta_i(p)$ is defined as (note that it differs from the $\delta_i(p)$ defined in~\cite{Arkani-Hamed:2020tuz}):
\begin{gather*}
\delta_i(p):=
\begin{cases}
\hphantom{-} 1 &{\rm if}~p=1 ~ {\rm mod}~ i ,\\
-1 &{\rm otherwise}.
\end{cases}
\end{gather*}
These point counts were previously performed using the $F$-polynomials of the cluster algebra. We find that the counting with the worldsheet variables yields the same result, as expected.

Substituting $p=1$, we get $55$ and $-674$ respectively, which agrees with the Euler characteristic, or the number of saddle points up to a sign (see~\eqref{numofsol}). For the $D_4$ space, we can also determine the dimensions of $1$-, $2$-, $3$- and $4$-cohomology to be exactly $16$, $93$, $231$ and $208$, which agree with (up to alternating signs) coefficients of $p^3$, $p^2$, $p^1$ and $p^0$ if we keep $\delta_3(p)=1$. This is of course consistent with the fact that the alternate sum of these numbers (including $1$ for $p^4$) should give the Euler characteristic (with $p=1$). We also remark that by substituting $p = -1$, we get $547$ and $6388$ respectively, which agree with the number of consistent sign patterns of the $u$-variables and the alphabet in terms of the $z$-variables.\footnote{The number of connected components may be counted by enumerating all possible sign patterns of the alphabet. The signs of the nonlinear $w_{i,j}$ letters that appear in the second line of~\eqref{Dnalphabet} may sometimes be determined by the relative order of the $\{z_i, z_j, z_-, z_+ \}$ variables. We may then exploit identities such as $w_{i,j} - w_{i,k} = (1+z_{i})(z_{j}-z_{k})$ and $w_{i,k} - w_{j,k} = (z_{i} - z_{j}) (-1 + z_{-} - z_{k} + z_{+})$ to eliminate inconsistent sign configurations of the $w_{i,j}$ letters.}
We find that the worldsheet variables allow us to perform numerical computations for higher-$D_n$ cases, which were not feasible using $F$-polynomials. For example, we can apply the finite-field method to the $D_6$, $D_7$ and $D_8$ cases, and by assuming that the result should be quasi-polynomials with $\delta_i(p)$ terms with $i<n$, we find the following formulas which exhibit a nice pattern
\begin{gather*}
|{\mathcal M}_{D_6}({\mathbb F}_p)| =p^6-36 p^5+530 p^4-4070 p^3+17140 p^2-37465 p+33301
\\
\hphantom{|{\mathcal M}_{D_6}({\mathbb F}_p)| =}{}+\delta_3(p) \big(15 p^2-246 p+987 \big)-\delta_4(p) (6 p-62 ) + 2 \delta_5(p),
\\
|{\mathcal M}_{D_7}({\mathbb F}_p)| = p^7-49 p^6+1014 p^5-11460 p^4+76215 p^3-297641 p^2+631280 p-562269
\\
\hphantom{|{\mathcal M}_{D_7}({\mathbb F}_p)| =}{}+\delta_3(p) \big(35 p^3-966 p^2+8736 p-25823 \big)-\delta_4(p) \big(21 p^2-476 p+2584 \big)
\\
\hphantom{|{\mathcal M}_{D_7}({\mathbb F}_p)| =}{} +\delta_5(p) (14 p-212 ) - 2 \delta_6(p),
 \\
 |{\mathcal M}_{D_8}({\mathbb F}_p)| =p^8-64 p^7+1771 p^6-27629 p^5+265335 p^4-1603427 p^3+5944309 p^2\\
\hphantom{|{\mathcal M}_{D_8}({\mathbb F}_p)| =}{}-12347924 p +11024276 +\delta_3(p) \big(70 p^4-2856 p^3+43092 p^2-284445 p \\
\hphantom{|{\mathcal M}_{D_8}({\mathbb F}_p)| =}{}+691145 \big)-\delta_4(p) \big(56 p^3-2072 p^2+24648 p-94824 \big)+\delta_5(p) \big(56 p^2-1808 p \\
\hphantom{|{\mathcal M}_{D_8}({\mathbb F}_p)| =}{} +13206 \big)-\delta _6(p) (16 p-368 ) + 3 \delta_7(p) .
\end{gather*}
 For $D_6$, we have generated such data for 37 prime numbers, from 7 up to 173; they are more than enough for us to find the first quasi-polynomial, which contains only 12 non-trivial coefficients. For $D_7$, we have generated such data of 21 prime numbers, from 7 to 89; they are more than enough to fix all 16 non-trivial coefficients in the second quasi-polynomial. Note that the coefficient of $\delta_6(p)$ can be absorbed into that of $\delta_3(p)$ since $\delta_6(p)=\delta_3(p)$, and the reason for splitting them in this way will become clear shortly. For $D_8$, we have generated the data of 21~prime numbers, from 11 to 97, they are just enough to fix all 21 non-trivial coefficients in the last quasi-polynomial. We emphasize that these are all conjectures about the point count which we do not know how to prove.

Substituting $p =1 $, we get $ 10 215$, $-183 256$ for $D_6$, $D_7$ respectively, which agrees with the Euler characteristic or numbers of saddle points (up to a sign)~\eqref{numofsol}. The agreement provides very strong support for these quasi-polynomials. For $D_8$, substituting $p = 1 $, we get $ 3 787 655$, which predicts that there should be 6 remaining solutions in addition to the 3787649 ones already found by HomotopyContinuation.jl~\eqref{numofsol}.

We have also conjectured quasi-polynomials of $D_9$ and $D_{10}$. For these cases, it has become more and more time-consuming to generate data for large prime numbers. For $D_9$, we have only generated data of 11 prime numbers, from 11 to 47 and for $D_{10}$, we have generated the data of 9 prime numbers, from 11 to 41. In neither case do we have enough data to completely fix the quasi-polynomials, but we can use some other data, e.g., the number of independent ${\rm d}\log$ $k$-forms for $k=1,2,3$, etc.\ (see below) to propose the following conjectures:
\begin{gather*}
|{\mathcal M}_{D_9}({\mathbb F}_p)| =p^9-81 p^8+2888 p^7-59416 p^6+776342 p^5- 6671938 p^4+ 37657424 p^3
 \\ \hphantom{|{\mathcal M}_{D_9}({\mathbb F}_p)| =}{} -134394689 p^2 + 274922647 p -245998148
 \\ \hphantom{|{\mathcal M}_{D_9}({\mathbb F}_p)| =}{} + \delta_3(p) \big(126 p^5- 7056 p^4 + 156240 p^3 - 1707489 p^2 + 9190566 p -19467332 \big)
 \\ \hphantom{|{\mathcal M}_{D_9}({\mathbb F}_p)| =}{} -\delta_4(p) \big(126 p^4-6720 p^3 + 130320 p^2 - 1094400p + 3365875 \big)
 \\
 \hphantom{|{\mathcal M}_{D_9}({\mathbb F}_p)| =}{} + \delta_5(p) \big(168 p^3-8640 p^2 + 135630 p - 670100 \big) \\
\hphantom{|{\mathcal M}_{D_9}({\mathbb F}_p)| =}{}
 -\delta _6(p) \big(72 p^2-3456p +33786 \big)
 + \delta_7(p) (27 p - 984 ) - 2 \delta_8(p),
 \\
|{\mathcal M}_{D_{10}}({\mathbb F}_p)| =
p^{10}-100 p^9+4464 p^8-117036 p^7+1993824 p^6-23038134 p^5
 \\
\hphantom{|{\mathcal M}_{D_{10}}({\mathbb F}_p)| =}{}+182628726 p^4 -979353469 p^3+3394906731 p^2-6862837036 p\\
\hphantom{|{\mathcal M}_{D_{10}}({\mathbb F}_p)| =}{}+6153401165 \\
\hphantom{|{\mathcal M}_{D_{10}}({\mathbb F}_p)| =}{} +\delta_3(p) \big(210 p^6-15372 p^5+464310 p^4-7398825 p^3+65494485 p^2 \\
\hphantom{|{\mathcal M}_{D_{10}}({\mathbb F}_p)| =}{} -304671969 p+582450185 \big)
 \\
\hphantom{|{\mathcal M}_{D_{10}}({\mathbb F}_p)| =}{}- \delta_4(p) \big(252 p^5-18060 p^4+504120 p^3-6878520 p^2+45974284 p \\
\hphantom{|{\mathcal M}_{D_{10}}({\mathbb F}_p)| =}{} -120501116 \big)
 \\
\hphantom{|{\mathcal M}_{D_{10}}({\mathbb F}_p)| =}{}+\delta_5(p) \big(420 p^4-30480 p^3+767070 p^2-8145380 p + 31203370 \big)
 \\
\hphantom{|{\mathcal M}_{D_{10}}({\mathbb F}_p)| =}{}- \delta _6(p) \big(240 p^3-18000 p^2+373140 p \big) + \delta_7(p) \big(135 p^2-10110 p+ 131289 \big)
\\ \hphantom{|{\mathcal M}_{D_{10}}({\mathbb F}_p)| =}{} - \delta_8(p) (20 p-1204 ) +3 \delta_9(p) .
\end{gather*}
Substituting $p = 1 $, we get $-88535634$ and $2308393321$, which predict the Euler characteristics (or the number of saddle points up to a sign) of $D_9$ and $D_{10}$ respectively.

It is computationally difficult to determine the number of saddle points (or equivalently the Euler characteristic) of the $D_n$ space for higher $n$. We just make a numerical observation: for $n=4,5,\dots, 10$, the number of saddle points can be approximated by $(1-n)^n \times \{0.679,0.658,0.654,0.655,0.657,0.660, 0.662\}$. It would be interesting to see if one could infer certain scaling behavior holds for large $n$, and if there might be some explanation for that.

Beyond $D_4$, we have not been able to compute dimensions of cohomology (due to the lack of efficient algorithms) except for the number of $1$-forms which is the number of letters, $n_1=n^2$. However, we conjecture that the dimension of the $k$-th cohomology is given by (the absolute value of) the coefficients of $p^{n{-}k}$ if we keep all $\delta_i(p)=1$. This conjecture is consistent with the observation that the Euler characteristic is obtained by plugging in $p=1$. To obtain some data that provides a lower bound on the dimensions, we can alternatively compute the number of independent ${\rm d}\log$ $k$-forms, $n_k$, obtained by wedging ${\rm d}\log$ of all the letters, for $k=1,2, \dots, n$. One needs to determine the rank of all the ${n^2 \choose k}$ ${\rm d}\log$ $k$-forms, by considering linear relations they satisfy. For higher $n$, it also becomes computationally difficult to determine $n_k$, but we have numerically determined such linear relations and obtained the number of independent $k$-forms for all $k$ up to $D_6$, and for $k=2,3$ up to $D_9$, see Table~\ref{table1}.
\begin{table}[t]\centering\renewcommand{\arraystretch}{1.2}
\caption{}\label{table1}
\vspace{1mm}

\begin{tabular}{c|c|c|c|c|c|c|c}
& 2-forms & 3-forms & 4-forms & 5-forms & 6-forms \\
 \hline
$D_4$ & 93 & 231 & 207 & --&--
 \\
$D_5$ & 244 & 1156 & 2649 & 2355&--
 \\
$D_6$ & 530 & 4070 & 17140 & 37465 & 33300 \\
$D_7$ & 1014 & 11460 & 76215 & &\\
$D_8$ & 1771 & 27629 & & &
\\
$D_9$ & 2888 & 59416 & & &
\\
$D_{10}$ & 4464 & & & &
\end{tabular}
\end{table}

Note that in addition to $n_1=n^2$, which is (the absolute value of) the coefficient of $p^{n{-}1}$, $n_2$ and~$n_3$, nicely match the next two coefficients of the quasi-polynomial, which are not affected by the correction terms, $\delta_i(p)$. As part of our conjecture above, these two numbers are conjectured to be the dimension of $2$- and $3$-cohomology, respectively, which seem to be generated by ${\rm d}\log$ forms (this is no longer true for $k\geq 4$). We propose that the number of independent (${\rm d}\log$) 2- and 3-forms are
\begin{gather*}
n_2 = \frac{1}{2} (n-1) \big(n^3-n+2\big) ,\qquad
n_3 = \frac{1}{3}\binom{n-1}{2} \big(n^4-3 n^2+5 n+3\big) .
\end{gather*}
As we will show in the next section, we can write down all the linear relations among the 2-forms, thus confirming the result for $n_2$. It would be interesting to find a similar argument for $n_3$.

\subsection{Cohomology of 2-forms}
The cohomology of the hyperplane arrangement complement is well understood as the Orlik--Solomon algebra~\cite{zbMATH03672449}. The linear relations among the ${\rm d}\log$ forms are the Orlik--Solomon ideals and the set of independent ${\rm d}\log$ forms form the Orlik--Solomon algebra. Our goal is to work out the analogues of the Orlik--Solomon ideals for nonlinear ${\rm d}\log$ forms.

For $D_{n}$, there are $\binom{n^2}{2}$ 2-forms of $\omega_{i,j} := {\rm d}\log z_{i,j}$ and $\chi_{i,j} := {\rm d}\log w_{i,j}$, including trivial ones: $\omega_{1,i} \wedge \omega_{2,i} = 0$.
The 2-forms of $\omega_{i,j}$ satisfy the Arnold relations
\[
\omega_{i,j} \wedge \omega_{j,k} + \omega_{j,k} \wedge \omega_{k,i} + \omega_{k,i} \wedge \omega_{i,j} = 0. 
\]
They provide $\binom{n}{3} + 2\binom{n}{2}$ constraints for $1\le i < j < k \le n$ and $1\le i < j \le n$, $k = n+2$ or $ n+3$.

Let
\begin{alignat*}{4}
&S^+_{i,j} = \omega _{i,n+3}+\omega _{j,n+2}, \qquad && S^-_{i,j} = \omega _{1,n+3} + \omega _{i,j}, &\\
&T^+_{i,j} = \omega_{1,i} + \omega_{j,n+2}, \qquad && T^-_{i,j} = \omega_{1,n+3} + \omega_{n+2,i}, & \\
&U^+_{i,j} = \omega_{i,n+2} +\omega_{j,n+3}, \qquad && U^-_{i,j} = \omega_{1,n+2} + \omega_{i,j} . &
\end{alignat*}
There are $3 \binom{n-1}{2}$ relations involving $\chi_{i,j}$ for $1 < i < j < n + 1$:
\begin{gather*}
S^+_{i,j} \wedge S^-_{i,j} + S^-_{i,j} \wedge \chi_{i,j} + \chi_{i,j} \wedge S^+_{i,j} = 0,\\
T^+_{i,j} \wedge T^-_{i,j} + T^-_{i,j} \wedge \chi_{i,j} + \chi_{i,j} \wedge T^+_{i,j} = 0,\\
U^+_{i,j} \wedge U^-_{i,j} + U^-_{i,j} \wedge \chi_{i,j} + \chi_{i,j} \wedge U^+_{i,j} = 0.
\end{gather*}
There are $2 \binom{n-1}{3}$ relations involving $\chi_{i,j}\wedge\chi_{i,k}$ and $\chi_{i,k} \wedge \chi_{j,k}$ for $1 < i < j < k < n + 1$:
\begin{gather*}
(\omega_{1,i} - \omega_{j,k}) \wedge \chi_{i,j} + \chi_{i,j}\wedge\chi_{i,k} + \chi_{i,k} \wedge (\omega_{1,i} - \omega_{j,k}) =0 , \\
 (\omega _{i,j}-\omega _{i,n+2}+\omega _{k,n+2}+\omega _{1,n+2} )\wedge\chi _{i,k}\ + \chi _{i,k}\wedge\chi _{j,k}\\
\quad{}+\chi _{j,k}\wedge (\omega _{i,j}-\omega _{j,n+2}+\omega _{k,n+2}+\omega _{1,n+2} )
- (\omega _{k,n+2}+\omega _{1,n+2} )\wedge (\omega _{i,n+2}-\omega _{j,n+2} ) = 0.
\end{gather*}
While we do not know if these are \emph{all} the cohomology relations among the two forms, they provide an upper bound on the number of independent 2-forms:
\begin{align*}
n_2 \le \binom{n^2}{2}-\left[\binom{n}{3} + 2\binom{n}{2}\right]-3 \binom{n-1}{2}-2 \binom{n-1}{3}
= \frac{1}{2} (n-1) \big(n^3-n+2\big) .
\end{align*}
It agrees with the prediction from the point count.

\section{Conclusion and discussions}

In this note, based on the connection between $Y$-systems and configuration spaces for finite-type cluster algebras, we have introduced worldsheet-like variables that generalize the worldsheet variables of the moduli space ${\mathcal M}_{0,n}$, and applied them for studying cluster configuration spaces. In particular, we have studied in detail these $z$-variables for type $D_n$, which have also provided a set of variables for the symbol alphabet of Feynman integrals as discussed in~\cite{Chicherin:2020umh, He:2021esx}. Our main results can be summarized as follows
\begin{itemize}\itemsep=0pt
 \item We provide a systematic derivation of the $u$-variables as generalized cross ratios of polynomials of $z$-variables. It is based on a gluing construction that defines an initial set of $u$-variables as cross ratios of a pair of $A_n$ worldsheets, and generates the remaining $u$-variables by the $Y$-system equations. We have characterized boundaries of the positive part of the space by degenerations of these polynomials. Upon fixing the gauge redundancy, such polynomials reproduce the alphabets for the $ABCD$ types in the literature and produce new alphabets for $E_6$, $F_4$, and $G_2$.
 \item We write the stringy canonical form and corresponding saddle-point equations in terms of these $z$-variables, which makes it easier to study their properties such as factorizations and pushforward. In particular, we find that the number of saddle points (or Euler characteristics) for the $D_n$ space up to $n=7$ (and those for the $E_6$, $F_4$, and $G_2$ spaces).
 \item We study topological properties of configuration spaces by counting the number of points in $\mathbb{F}_p$: we conjecture that the point count is a quasi-polynomial with specific ``correction terms'' for type $D_n$, based on empirical evidence up to $n=10$, which predicts dimensions of cohomology and the Euler characteristic, etc.
\end{itemize}

There are numerous questions for future investigations raised by our preliminary studies. First of all, we would like to understand possible ``physical'' meanings of these worldsheet-like variables. For cluster configuration spaces, this may be related to the possible meaning of cluster string integrals: for the $D_n$ case, the stringy integral has not only the correct $\alpha'\to 0$ limit as the one-loop planar $\phi^3$ integrand, but also the correct factorization at massless poles for finite~$\alpha'$; it would be highly desirable if we can understand better such stringy integrals using these $z$-variables. As we have seen in~\cite{Chicherin:2020umh, He:2021esx, He:2021non}, the letters of $D_n$ using $z$-variables (for $n\leq 6$) have nicely appeared in the symbol of ``ladder-type'' Feynman integrals to all loops, where the $z$'s are nothing but all the {\it last entries}. This has opened up another direction for studying possible physical meanings of such worldsheet-like variables.

Moreover, it would be interesting to see if one can prove the quasi-polynomials for the point count of $D_n$ (up to $n=10$) and maybe even find a general pattern. While they do not correspond to hyperplane arrangement complement, the $z$-variables make it clear that we only need to deal with quadratic polynomials $w_{i,j}$, in addition to those linear factors. We have not been able to come up with any conjecture for the point count for the $E_6$ or $F_4$ case, which seems to require at least some new forms of ``correction terms". It would be highly desirable to see what can we learn about cluster algebras from these topological properties of the corresponding cluster configuration spaces.

Note that the type-$A$ cluster algebra is equivalent to the cluster structure on the Grassmannian of 2-planes, both of which can be used to describe moduli space for string theory and for the Cachazo--He--Yuan formula~\cite{Cachazo:2013hca}.
Other finite-type cluster algebras and higher-$k$ Grassmannians generalize them in two different ways. Higher-$k$ Grassmannian stringy integrals have been studied in~\cite{Arkani-Hamed:2019mrd,He:2020ray}, whose leading orders are equivalent to the higher-$k$ CHY formulas (or CEGM generalized bi-adjoint amplitudes)~\cite{Abhishek:2020xfy,Cachazo:2019ngv}. Tropical Grassmannian~\cite{Drummond:2019qjk,Drummond:2020kqg,Henke:2019hve,Henke:2021avn,Lukowski:2020dpn}, matroid subdivisions~\cite{Cachazo:2020uup,Cachazo:2020wgu,Early:2019eun,Early:2019zyi}, and the planar collections of Feynman diagrams~\cite{Borges:2019csl,Cachazo:2019xjx,Guevara:2020lek} are also found to be very useful to study them. Compared to higher-$k$ Grassmannians, factorization behaviors of finite-type cluster algebras are better understood. As the worldsheet variables have been discovered for other finite-type cluster algebras, it is natural to compare our study of cluster configuration spaces to those Grassmannian cases. It would be interesting to compare the topological properties of their configuration spaces including the Euler characteristic~\cite{Agostini:2021rze,Cachazo:2019apa,Cachazo:2019ble,Skorobogatov_1996,Sturmfels:2020mpv}), their ABHY realizations, and even applications to the symbol alphabets that appeared in higher-loop integrals of SYM, etc.~\cite{Arkani-Hamed:2019rds,Drummond:2019cxm,He:2021non,Ren:2021ztg}.

Last but not least, an important question already mentioned in~\cite{Arkani-Hamed:2020tuz} is if we could classify all the connected components of the $D_n$ configuration space and find their corresponding canonical forms, which have not been understood even for $D_4$. It becomes much more viable now with these $z$-variables, and this would enable us to understand better the cohomologies and more importantly the extension of the stringy integrals (so far for the positive part) to general ``off-diagonal'' real and complex integrals~\cite{Arkani-Hamed:2019mrd}.

\subsection*{Acknowledgements}
We thank Zhenjie Li, Qinglin Yang, and Yichao Tang for collaborations on related projects, and Yang Zhang for help with the USTC computing platform. YZ would like to thank Alex Edison for the discussions on the number of independent forms. PZ would like to thank Yuqi Li for the discussions on the string worldsheet. We thank the anonymous referees for detailed comments that help improve the presentation of the draft. SH's research was supported in part by the National Natural Science Foundation of China under Grant No.\ 11935013, 11947301, 12047502, 12047503, 12247103, 12225510. The research of YZ is supported by the Knut and Alice Wallenberg Foundation under the grant KAW 2018.0116: From Scattering Amplitudes to Gravitational Waves. PZ is supported by a~China Postdoctoral Science Foundation Special Fund, grant number Y9Y2231.

\pdfbookmark[1]{References}{ref}
\LastPageEnding

\end{document}